\documentclass[a4paper,fleqn,usenatbib]{mnras}

\usepackage{newtxtext,newtxmath}

\usepackage[T1]{fontenc}
\usepackage{ae,aecompl}

\usepackage{graphicx}	
\usepackage{amsmath}	
\usepackage{rotating}

\usepackage{color,soul}
\definecolor{lightblue}{rgb}{.70,.95,1}
\sethlcolor{lightblue}

\newcommand{\Eexc}{$E_{\rm exc}$}
\newcommand{\Teff}{T_{\rm eff}}
\newcommand{\logg}{\log {g}}
\newcommand{\kms}{km\,s$^{-1}$}
\newcommand{\eps}[1]{\log\varepsilon_{\rm #1}}
\def\ione{\,{\sc i}}
\def\ii{\,{\sc ii}}
\newcommand{\feh}{\rm{[Fe/H]}}
\newcommand{\cfe}{\rm{[C/Fe]}}

\title[PIGS VII: first CEMP-r/s star in the bulge]{The Pristine Inner Galaxy Survey (PIGS) VII: a discovery of the first inner Galaxy CEMP-r/s star\thanks{Based on observations made with the Very Large Telescope (VLT)}}

\author[L. Mashonkina et al.]{
L.~Mashonkina,$^{1}$\thanks{E-mail: lima@inasan.ru}
A.~Arentsen,$^{2}$
D.~S.~Aguado,$^{3,4}$
 A.~Smogorzhevskii,$^{1,5}$
M.~Hampel,$^{6}$
\newauthor A.I~Karakas,$^{6,7}$
F.~Sestito,$^{8}$ 
N.~F.~Martin,$^{9,10}$
K.~A.~Venn,$^{8}$
J.~I.~Gonz{\'a}lez Hern{\'a}ndez$^{3,4}$ \\
$^{1}$ Institute of Astronomy of the Russian Academy of Sciences, Pyatnitskaya st. 48, 119017, Moscow, Russia \\
$^{2}$ Institute of Astronomy, University of Cambridge, Madingley Road, Cambridge CB3 0HA, UK \\
$^{3}$Instituto de Astrof\'{\i}sica de Canarias,
              V\'{\i}a L\'actea, 38205 La Laguna, Tenerife, Spain\\
$^{4}$Universidad de La Laguna, Departamento de Astrof\'{\i}sica, 
             38206 La Laguna, Tenerife, Spain \\
$^{5}$ M. V. Lomonosov Moscow State University, Kolmogorova st. 1, 119991, Moscow, Russia \\
$^{6}$ School of Physics \& Astronomy, Monash University, Clayton, VIC 3800, Australia \\
$^{7}$ ARC Centre of Excellence for All Sky Astrophysics in 3 Dimensions (ASTRO 3D), Melbourne, VIC, Australia \\
$^{8}$ Department of Physics and Astronomy, University of Victoria, PO Box 3055, STN CSC, Victoria BC V8W 3P6, Canada \\
$^{9}$ Observatoire Astronomique de Strasbourg, Universit\'e de Strasbourg, CNRS, UMR 7550, F-67000 Strasbourg, France \\
$^{10}$ Max-Planck-Institut f\"{u}r Astronomie, K\"{o}nigstuhl 17, D-69117 Heidelberg, Germany
}

\date{Accepted XXX. Received YYY; in original form ZZZ}

\pubyear{2023}

\begin{document}
\label{firstpage}
\pagerange{\pageref{firstpage}--\pageref{lastpage}}
\maketitle

\begin{abstract}
Well-studied very metal-poor (VMP, [Fe/H] $< -2$) stars in the inner Galaxy are few in number, and they are of special interest because they are expected to be among the oldest stars in the Milky Way. We present high-resolution spectroscopic follow-up of the carbon-enhanced metal-poor (CEMP) star Pristine\_184237.56-260624.5 (hereafter Pr184237) identified in the Pristine Inner Galaxy Survey. This star has an apocentre of $\sim 2.6$~kpc. Its atmospheric parameters ($\Teff$ = 5100~K, $\logg$ = 2.0, [Fe/H] = $-2.60$) were derived based on the non-local thermodynamic equilibrium (NLTE) line formation. We determined abundances for 32 elements, including 15 heavy elements beyond the iron group. The NLTE abundances were calculated for 13 elements from Na to Pb. Pr184237 is strongly enhanced in C, N, O, and both s- and r-process elements from Ba to Pb; it reveals a low carbon isotope ratio of $^{12}$C/$^{13}$C = 7. The element abundance pattern in the Na-Zn range is typical of halo stars. With [Ba/Eu] = 0.32, Pr184237 is the first star of the CEMP-r/s subclass identified in the inner Galaxy. Variations in radial velocity suggest binarity. We tested whether a pollution by the s- or i-process material produced in the more massive and evolved companion can form the observed abundance pattern and find that an i-process in the asymptotic giant branch star with a progenitor mass of 1.0-2.0$M_{\odot}$ can be the solution. 
\end{abstract}



\section{Introduction}

Studies of very metal-poor (VMP, [Fe/H]\footnote{In the classical notation, where [X/Y] = $\log(N_{\rm X}/N_{\rm Y})_{star} - \log(N_{\rm X}/N_{\rm Y})_{\odot}$ for each pair of elements X and Y.} $< -2.0$) stellar populations in the Milky Way are important to understand the early Universe. Their detailed chemical abundances teach us about the properties of the First Stars and early star formation, and their chemo-dynamics shed light on the early formation history of our Galaxy \citep{frebelnorrisreview}. The central regions of our Galaxy ($\lesssim 5$~kpc) are predicted to host the oldest metal-poor stars \citep{2010ApJ...708.1398T}, which are important probes of the earliest metal-free stars in the Universe. In recent years, significant efforts have been made to build larger samples of VMP stars in the inner Galaxy \citep{2015Natur.527..484H,2016MNRAS.460..884H,2019MNRAS.488.2283L,2020MNRAS.496.4964A} -- a challenging endeavour because the overwhelming majority of bulge stars have high metallicity, and the high extinction and the relatively large distance from the Sun make clean selections of metal-poor stars difficult. These surveys typically employ narrow-band photometry to identify metal-poor stars, which has been very effective.

Recent studies show that VMP stars in the inner Galaxy appear to have similar abundances to stars in previous halo samples, although there are some subtle differences in, e.g., the scatter and correlations of various abundances \citep{2016MNRAS.460..884H,2016A&A...587A.124K,2019MNRAS.488.2283L,2023MNRAS.518.4557S}. One other striking difference is an apparent lack of carbon-enhanced metal-poor (CEMP) stars \citep{2016MNRAS.460..884H, 2021MNRAS.505.1239A}, with the common definition for CEMP stars having [C/Fe] $>+0.7$ \citep{2007ApJ...655..492A}. In the Galactic halo, many of the most metal-poor stars are found to be rich in carbon, representing 20\%\ and 43\%\ of stars with [Fe/H] $\le -2$ and $\le -3$, respectively \citep{2014ApJ...797...21P}.
For the bulge, \citet{2021MNRAS.505.1239A} found a CEMP fraction of only 16\%\ for [Fe/H] $< -2.5$ in the low-resolution spectroscopic Pristine Inner Galaxy Survey (PIGS). They discuss that it might partly be related to selection effects against carbon-rich stars, but argue that it cannot be explained away entirely. The lower CEMP fraction could be due to a lower number of binary stars in the inner Galaxy, and/or due to faster and more intense early star formation -- both scenarios would be indications that the metal-poor stellar populations in the inner Galaxy and the more distant halo formed in different environments.

Depending on abundances of the elements produced presumably in the slow (s) or rapid (r) neutron-capture processes, which can be derived from high-resolution spectroscopy, CEMP stars are separated into CEMP-r ([Eu/Fe] $> 1$), CEMP-s ([Ba/Fe] $> 1$ and [Ba/Eu] $> 0.5$), CEMP-r/s (enhanced in both barium and europium with $0 <$ [Ba/Eu] $< 0.5$), and CEMP-no ([Ba/Fe] $< 0$) stars, as suggested by \citet{2005ARA&A..43..531B}. Slightly different numbers are recommended by \citet{2016A&A...587A..50A} and \citet{2018ARNPS..68..237F} to characterise different groups. Being purely phenomenological, this separation turns out to be indicative of the nature of stars in different groups. Excess of carbon and s-process elements in a CEMP-s star is thought to be the result of mass transfer from an asymptotic giant branch (AGB) star to a companion in a binary system. This hypothesis is supported by radial velocity variations observed for many CEMP-s stars \citep{2016A&A...588A...3H}. The CEMP-no stars, which are typically not in binary systems, likely formed out of the interstellar matter enriched by the Population~III stars, which either were rapidly rotating \citep{2006A&A...449L..27C} or exploded as faint supernovae \citep[][and references therein]{2013ARA&A..51..457N}. CEMP-r stars are similar to CEMP-no stars, except that they are enhanced in r-process elements, which they were likely born with as well. The nature of CEMP-r/s stars is under debate, and various scenarios of enrichment in heavy elements are considered, from mass transfer in a binary system formed out of the interstellar matter rich in the r-process elements to the operation of the intermediate neutron-capture process \citep[i-process, see][for a review]{2021A&A...649A..49G}.





Detailed studies of CEMP stars can provide a unique information about the early chemical enrichment of the inner Galaxy, but they are only available in the literature for a very few stars.
Four CEMP-no stars, all with [Fe/H] $< -3$ and low Ba abundances, were discovered in the EMBLA (Extremely Metal-poor BuLge stars with AAOmega) sample \citep{2015Natur.527..484H,2016MNRAS.460..884H}: one was reported by \citet{2015Natur.527..484H}, and three additional stars were found by \citet{2021MNRAS.505.1239A} when applying the evolutionary carbon corrections from \citet{2014ApJ...797...21P}.
The three CEMP stars with close together metallicities of [Fe/H] $\simeq -2.5$ were found to be strongly enhanced in Ba, with [Ba/Fe] $> +1$. One of them was classified by \citet{2016A&A...587A.124K} as CEMP-s based on low upper limits for abundances of the r-process elements Eu and Dy. None of the r-process elements were measured by \citet{2023MNRAS.518.4557S} in the remaining two CEMP stars, due to a limited wavelength range of their Gemini GRACES spectra.

In this work, we report the detailed analysis of a new CEMP star in the inner Galaxy, Pristine\_184237.56-260624.5 (for short, Pr184237), with [Fe/H] = $-2.6$ and [C/Fe] = $+1.8$, which is enhanced in both s- and r-process elements. It was observed as part of a high-resolution spectroscopic follow-up campaign of the most metal-poor stars in PIGS, undertaken with UVES\footnote{Ultraviolet and Visual Echelle Spectrograph} at the VLT. It deserved a careful, dedicated analysis due to the high carbon and neutron-capture element abundances, which we report on here separately from the main sample. We performed a detailed line-by-line analysis of the spectrum of Pr184237 in a wide wavelength range and determined abundances for 32 chemical elements, including 15 neutron-capture elements. Abundances for 13 chemical elements are based on the non-local thermodynamic equilibrium (NLTE) line formation, which is important because the physical conditions in atmospheres of VMP giants are favorable for the departures from local thermodynamic equilibrium (LTE).

The paper is organised as follows. We introduce PIGS and the high-resolution observations of Pr184237 in Sect.~\ref{sect:obs}. Section~\ref{sect:method} describes the NLTE methods and codes used. Section~\ref{sect:atmos} discusses the determination of the star's atmospheric parameters.
Elemental abundances are derived in Sect.~\ref{Sect:abund}, and Sect.~\ref{Sect:origin}  discusses the possible origins of the heavy element abundance pattern of Pr184237.
Our conclusions are given in Sect.~\ref{Sect:conclusion}.

\section{Observations of Pr184237}\label{sect:obs}

The Pristine Inner Galaxy Survey (PIGS), an extension of the main Pristine survey \citep[see][for an overview]{2017MNRAS.471.2587S}, was initiated to build an unprecedentedly large sample of metal-poor and very metal-poor stars for chemo-dynamical studies of the inner regions of the Galaxy \citep{2020MNRAS.491L..11A,2020MNRAS.496.4964A,2021MNRAS.505.1239A}.
Using metallicity-sensitive $CaHK$ photometry from the Canada-France-Hawaii-Telescope (CFHT), PIGS selects metal-poor candidates for spectroscopic follow-up. The next step is an intermediate-resolution spectroscopy with the AAOmega spectrograph \citep{2004SPIE.5492..389S} on the Anglo-Australian Telescope combined with determinations of the stellar atmosphere parameters using the ULySS \citep{2009A&A...501.1269K} and FERRE \citep{2006ApJ...636..804A} tools. A total of 1900 VMP stars were identified in the PIGS-AAT sample. Aiming to understand the nucleosynthesis processes in the ancient inner Galaxy, $\sim$20 of the most metal-poor stars were selected from the AAT sample for high-resolution spectroscopy with the UVES/VLT, one of which was Pr184237.

Pr184237 was found to have [Fe/H] = $-2.8$ and [C/Fe] = $+2.0$ from the low-resolution spectroscopy and was recognised as a CEMP star in \citet{2021MNRAS.505.1239A}. The AAT barycentric radial velocity is $-12.5 \pm 2.0$ \kms. It is located at (RA,\,Dec) = (18:42:37.56,$-$26:06:24.5) and $(l,b) = (8.514105^\circ,-9.783936^\circ)$ and identified with Gaia DR3 source\_id 4073253907337129472 of G magnitude = 15.52 \citep{gaiadr3}. It has E(B-V) = 0.38 \citep{green19}.

\subsection{UVES observations}

The UVES observations of Pr184237 were originally scheduled for period 105 in queue mode. However, due to COVID restrictions in Paranal (Chile), observations were taken in period 107. The employed setup was $\#Dic1\,(390+580)$ with a $1\farcs2$ slit, $1\times1$ binning and low readout speed. This configuration led to a resolving power of R$\sim45,000$ in the blue part of the spectrum ($330-452$~{nm}) and R$\sim41,500$ for the red ($480-680$~{nm}). 
Three observations were taken between the 3rd and the 4th of June 2021 with an exposure time of 3000\,s each and an average signal-to-noise ratio (SNR) of 7 and 32 at 380 and 510\,nm, respectively.

The data were reduced by the ESO pipeline and retrieved by a query on the phase 3 online interface\footnote{\url{https://www.eso.org/sci/observing/phase3.html}}. The barycentric correction was applied with IRAF and the radial velocity was calculated with a cross correlation function to a UVES template with the {\tt fxcor} package, after which the three spectra were combined in the rest frame with a ``sigclip'' algorithm. The three radial velocities are reported in Table~\ref{Tab:properties}, together with the AAT radial velocity.

A first full-spectrum fitting analysis of the UVES spectrum was performed with the {\tt FERRE}\footnote{\tt http://github.com/callendeprieto/ferre} code following the same procedure explained in \citet{aguado21a, aguado21b}. In brief, the code is able to fit the data and interpolate between the nodes of a grid of stellar models computed with ASSET \citep{koe08} and ATLAS \citep{sbo05} codes of spectral synthesis. These models already contain carbon-enrichment, an important feature when analyzing metal-poor stars \citep[see e.g.][for more information about the carbon models]{agu17I}. The four parameters {\tt FERRE} derived for Pr184237 are effective temperature ($\Teff=5448\pm123$\,K), surface gravity ($\logg=2.01\pm0.26$), overall metallicity ($\feh=-2.46\pm0.10$), and carbon enrichment ($\cfe=2.19\pm0.12$). A preliminary automatic abundance analysis with FERRE indicated that Pr184237 is enhanced in both s-process ([Ba/Fe] = 1.04, [La/Fe] = 0.56) and r-process ([Eu/Fe] = 0.6) elements, suggesting that this is the first CEMP-r/s star in the inner Galaxy. This encouraged us to do a dedicated detailed analysis (see Sect.~\ref{sect:atmos}-\ref{Sect:abund}). The FERRE analysis for the full PIGS-UVES sample will be presented in a forthcoming paper (Aguado et al., in prep.).  

\begin{table} 
 \centering
 \caption{\label{Tab:properties} Radial velocities of Pr184237.  
 }
  \begin{tabular}{lccc}
   \hline\hline
   Instrument & MJD  & RV \\
    & [days] &  [\kms] \\
   \hline
   AAT  & 58337.44 &  -12.5 $\pm$ 2.0 \\
   UVES & 59368.37 &  -39.3 $\pm$ 1.0 \\
   UVES & 59369.33 &  -39.1 $\pm$ 1.0 \\
   UVES & 59369.36 &  -39.8 $\pm$ 1.0 \\
   \hline
  \end{tabular}
\end{table}

\subsection{Variability and binarity}

The AAT baricentric radial velocity ($-12.5 \pm 2.0$~\kms) is significantly different from the three UVES radial velocities taken almost three years later ($-39.4$~\kms\ with a standard deviation of 0.3~\kms). The precision of the AAT radial velocities was estimated by \citet{2020MNRAS.496.4964A} to be around 2~\kms. A more recent comparison with Gaia DR3 radial velocities for the full AAT sample (where available, only for brighter stars and not for Pr184237) shows that the uncertainty estimate was correct and possibly even slightly over-estimated (Arentsen et al. in prep). Therefore the difference between the AAT and UVES radial velocities can be taken as a clear indication that this star is in a binary (or multiple) system. 

We also found that Pr184237 was flagged as a photometrically variable star in Gaia DR3, with the classification "ELL" \citep[best class score = 0.58]{gaiadr3variable} -- an ellipsoidal variable, which is a close binary system where the stars are deforming and projection effects during the orbit cause photometric variability. There is a Gaia light curve available for this star, with variations on the order of $\sim 0.03$~mag on a short timescale (less than a few days). Fitting the epoch photometry with {\sc lightkurve} \citep{lightkurve} we found a best period of 1.8 days, but it was not very significant. Such a short period (if this star really was an ellipsoidal variable in a binary system) is not consistent with our UVES observations. The UVES radial velocities taken over two days do not show any significant radial velocity variation, and there is also no sign of severe rotational line broadening. One could maybe imagine a scenario where this star is in a triple system, orbiting in a longer period around an inner short-period binary. In this scenario the short-term variability could come from the inner binary. If the close binary consists of two white dwarfs, they should be too faint to contribute to the photometric variability since a white dwarf is $\sim 10$ mag fainter than an RGB star like Pr184237 (related to the rapidly accreting white dwarf scenario discussed in Section~\ref{sec:iprocess}). 

However, inspecting images of Pr184237, we found that there is a faint star ($\sim 4$ mag fainter, G = 19.9) within $\sim 2''$ of the main target. It has different proper motions so is not associated with Pr184237, but it might affect the photometry. We checked the Gaia scanning law and found that, for the epochs where Pr184237 appears brighter, Gaia scanned the sky exactly in the direction in which the two stars are aligned, and, when Pr184237 appears fainter, it was scanned perpendicular to that axis. The difference in flux between the two stars is similar to the amplitude of the photometric variability. Combined with the lack of evidence for a short orbital period from our UVES observations, we therefore concluded that the Gaia photometric variability is most likely spurious and not related to Pr184237 itself.

\subsection{Orbit in the Milky Way}

Stars that are currently in the inner Galaxy do not necessarily stay in the inner Galaxy -- they could be outer halo stars just moving through. To test this for Pr184237, we derived its distance and orbital properties. The distance and orbits were derived adopting the same methodology used by \citet{2023MNRAS.518.4557S} for the PIGS follow-up with the GRACES high-resolution spectrograph. Briefly, the Bayesian distance derivation uses the Gaia DR3 parallax and its uncertainty plus a prior on the stellar density distribution containing a disc and a halo (see \citealt{sestito19} for more details about the method). For Pr184237, it resulted in a median distance of $7.97 \pm 2.57$~kpc (corresponding to a distance from the Galactic centre $d_\mathrm{GC} \approx 1.8$~kpc).
The orbital integration is performed with GALPY \citep{2015ApJS..216...29B} with the inclusion of a rotating bar in the gravitational potential. Then the orbital parameters are calculated from the distance probability distribution function. More details can be found in 
\citet{2023MNRAS.518.4557S}. We adopt a radial velocity of $-26.0$~\kms, the average between the AAT and the UVES epochs, with an inflated uncertainty of 10~\kms\ to represent that we do not actually know the systemic velocity.
Adopting the above distance and radial velocity with their uncertainties and combining it with the Gaia DR3 proper motions, we derive the orbital properties of  Pr184237, and find it has a median apocentre of $2.56 \pm 0.93$~kpc with a pericentre less than 1~kpc -- it appears to be a real inner Galaxy star and not just a halo star passing through. Given that it is so tightly confined, the star is most likely to be of a ``proto-Galactic'' origin \citep{rix22,belokurov22} rather than accreted from a dwarf galaxy, or if it was accreted, this must have happened very early on.


\section{Analysis method}\label{sect:method}

We determined the NLTE abundances for Na, Mg, Al, Si, Ca, Sc, Ti, Fe, Zn, Sr, Ba, Eu, and Pb. Hereafter, these elements are referred to as NLTE species.
For individual lines of Ti\ii, Fe\ione, and Zn\ione, the NLTE abundances were computed by adding the NLTE abundance corrections, $\Delta_{\rm NLTE} = \eps{\rm NLTE} - \eps{\rm LTE}$, to the LTE abundances. The $\Delta_{\rm NLTE}$ grids and interpolation tools are provided by \citet[][Ti\ii, Fe\ione]{Mashonkina_dnlte2016} and \citet[][Zn\ione]{2022MNRAS.515.1510S}.
For the remaining NLTE species, we performed the NLTE calculations using individual atmospheric parameters and elemental abundances of Pr184237 and the NLTE methods developed in our previous studies for Na\ione\ \citep{alexeeva_na}, Mg\ione\ \citep{mash_mg13}, Al\ione\ \citep{mash_al2016}, Si\ione -Si\ii\ \citep{2020MNRAS.493.6095M}, Ca\ione -Ca\ii\ \citep{2017AA...605A..53M}, Sc\ii\ \citep{nlte_sc2}, Sr\ii\ \citep{2022MNRAS.509.3626M}, Ba\ii\ \citep{2019AstL...45..341M}, Eu\ii\ \citep{mash_eu}, and Pb\ione\ \citep{Mashonkina_pb}.


The coupled radiative transfer and statistical equilibrium (SE) equations were solved with the code {\sc detail} \citep{Butler84,Giddings81}, where the background opacity was updated as described by
\citet{mash_fe}. The computed departure coefficients, $b_i = n_i^{\rm NLTE}/n_i^{\rm LTE}$, were then used by the code {\sc synthV}\_NLTE \citep{2019ASPC} to calculate the NLTE line profiles for a given NLTE species. Here, $n_i^{\rm NLTE}$ and $n_i^{\rm LTE}$ are the SE and thermal (Saha-Boltzmann) number densities, respectively.

The LTE and NLTE abundances were derived from line profile fitting using the code {\sc synthV}\_NLTE \citep{2019ASPC}. The best fit to the observed spectrum is obtained automatically using the {\sc IDL binmag} code by \citet{2018ascl.soft05015K}.
The line list required for calculations of the synthetic spectra, together with atomic data, were taken from the most recent version of Vienna Atomic Line Database \citep[VALD,][]{2015PhyS...90e4005R}
that includes isotopic splitting (IS) and hyper-fine splitting (HFS) components where they are available \citep{2019ARep...63.1010P}. The lines of the NLTE species were computed by implementing the departure coefficients from {\sc detail}, while LTE was assumed for the lines of the other species.

We used the homogeneous spherical model atmospheres with standard abundances \citep{Gustafssonetal:2008} from the MARCS website\footnote{\url{https://marcs.astro.uu.se}}. They were interpolated at the given  effective temperature ($\Teff$), surface gravity ($\logg$), and metallicity ([Fe/H]), applying the interpolation routine written by Thomas Masseron and available on the same website.

\section{Atmospheric parameters}\label{sect:atmos}

For Pr184237, the preliminary automatic FERRE analysis (Aguado et al., in prep.) yielded $\Teff$ = 5450$\pm123$~K, $\logg$ = 2.01$\pm0.26$, and [Fe/H] = $-2.46\pm0.10$.
In view of the extreme importance of derived elemental abundances for understanding the phenomenon of a CEMP-r/s star and for learning mechanisms of the early chemical enrichment of the inner Galaxy, we double checked and improved the atmospheric parameters of this star.

\begin{figure}  
 \begin{minipage}{85mm}
\centering
	\includegraphics[width=0.99\textwidth, clip]{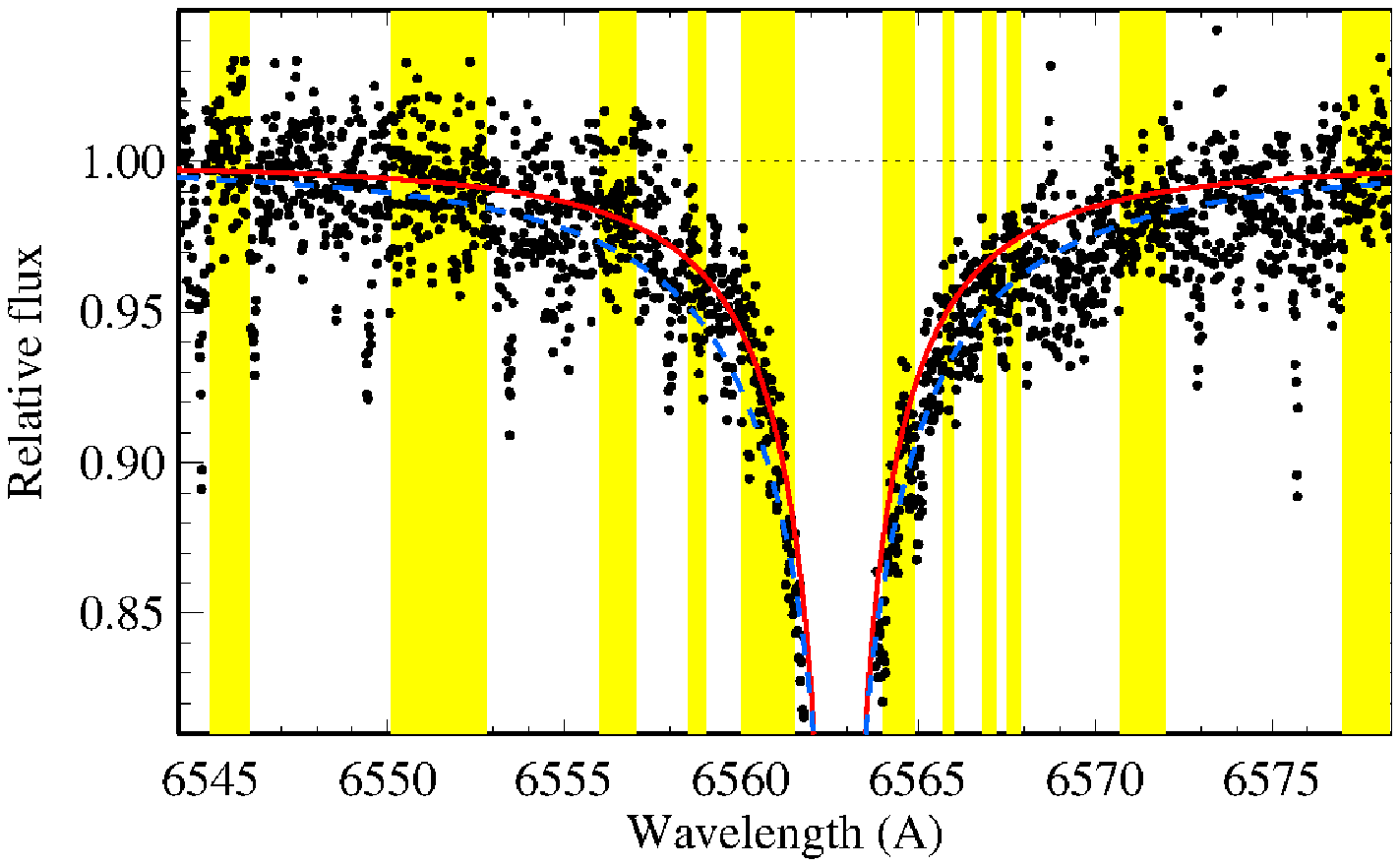}
	\includegraphics[width=0.99\textwidth, clip]{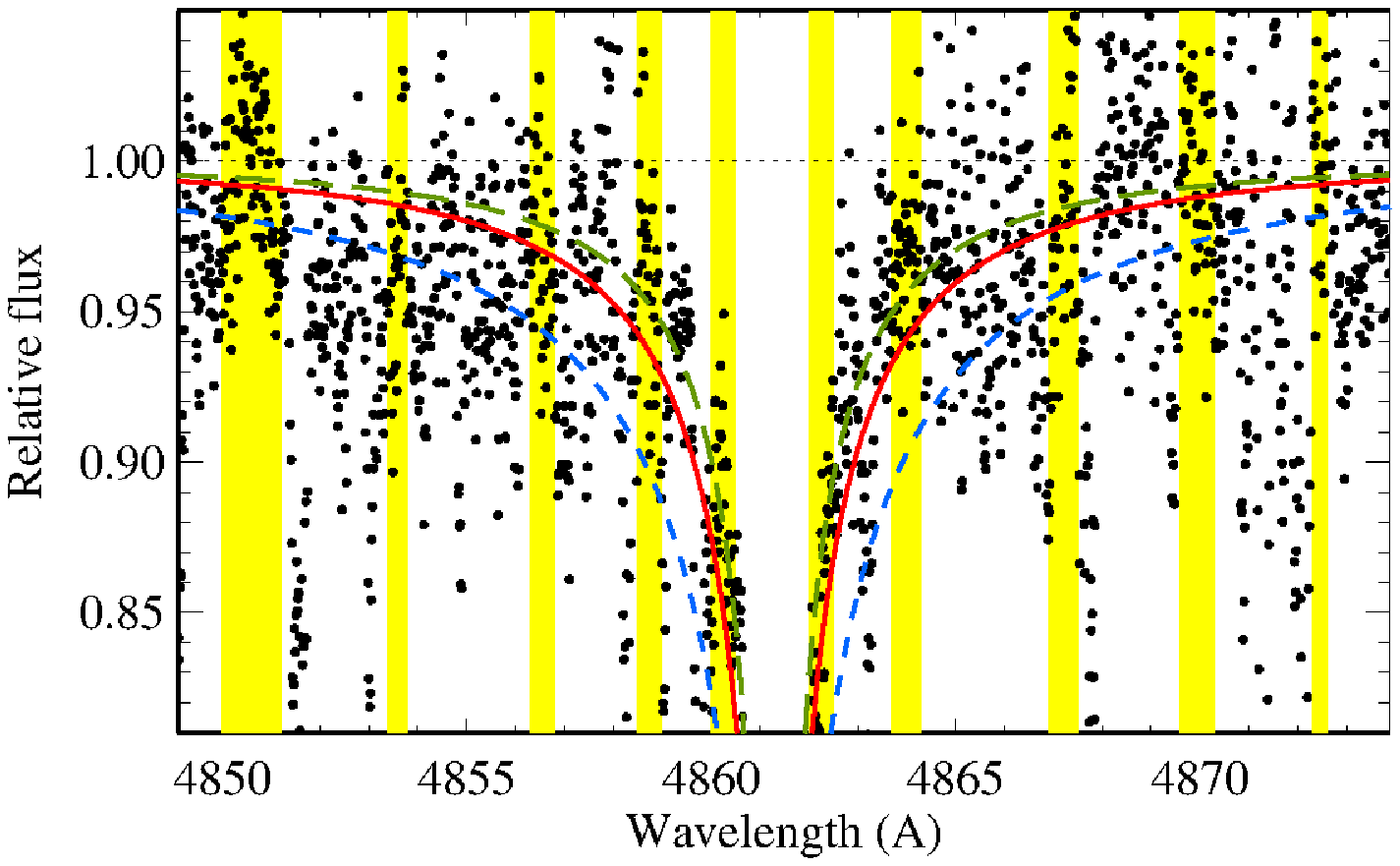}
  \caption{\label{Fig:hyd} H$_\alpha$ (top panel) and H$_\beta$ (bottom panel) line profiles in Pr184237 (bold dots) compared with the theoretical 3D-NLTE profiles from \citet{2018A&A...615A.139A} for $\Teff$ = 5100~K (red continuous curve), $\Teff$ = 5450~K (blue dashed curve), and $\Teff$ = 4900~K (green long-dashed curve, for H$_\beta$ only). Everywhere, $\logg$ = 2.0 and [Fe/H] = $-2.5$. The yellow shaded regions show the spectral windows used for the fit. }
\end{minipage}
\end{figure}

The effective temperature is derived by fitting the theoretical profiles computed by \citet{2018A&A...615A.139A} with three-dimensional (3D) hydrodynamic model atmospheres and based on the NLTE line formation to the observed wavelength regions free of blending lines in the wings of the Balmer H$_\alpha$ and H$_\beta$ lines (Fig.~\ref{Fig:hyd}). For both lines, the theoretical profiles corresponding to $\Teff$ = 5450~K are apparently deeper compared with the observed ones. Using a $\chi^2$ minimization, we find that the best fit to H$_\alpha$ is achieved with $\Teff$ = 5100~K ($\chi^2$ = 0.127). For H$_\beta$, equal preference can be given to $\Teff$ = 5100~K and 4900~K due to similar $\chi^2 \simeq$ 0.4. We give the greater weight to $\Teff$ from H$_\alpha$ because of less important blending effects and higher S/N, and adopt $\Teff$ = 5100~K as our final effective temperature. From a comparison of the results from two lines,
we estimate the uncertainty in $\Teff$ as 100~K.

The surface gravity is constrained from a requirement of equal NLTE abundances determined from lines of the two ionization stages, \ion{Fe}{i} and \ion{Fe}{ii}. We use the \ion{Fe}{i} and \ion{Fe}{ii} lines and their atomic parameters, carefully selected and checked in our previous studies of VMP stars \citep{dsph_parameters}. In brief, for \ion{Fe}{i}, we use presumably experimental $gf$-values, as listed by VALD, while, for \ion{Fe}{ii}, we apply $gf$-values from \citet{RU} that were corrected by $+0.11$~dex, following the recommendation of \citet{Grevesse1999}. The exceptions are \ion{Fe}{ii} 4923 and 5018\,\AA, for which $gf$-values were obtained by averaging the data from the four sources --  \citet{Bridges_fe2,1983A&AS...52...37M}, \citet{RU}, and \citet{MB09}. The van der Waals damping constants were taken from VALD.

\begin{table} 
 \centering
 \caption{\label{Tab:linelist} Line atomic data, LTE, and NLTE abundances, $\eps{}$, for individual lines in Pr184237.} 
 \begin{tabular}{lccrcc}
\hline\hline \noalign{\smallskip}
Species & $\lambda$ & \Eexc & log~$gf$ &  LTE & NLTE \\
        & [\AA]     & [eV]  &       &      &      \\
\noalign{\smallskip} \hline \noalign{\smallskip}
$^{12}$CH      &  4210.94 & 0.46 & -1.34 &  7.49$^1$ & \\
$^{12}$CH      &  4210.99 & 0.46 & -1.32 &  7.49$^1$ & \\
$^{13}$CH      &  4211.44 & 0.46 & -1.34 &  7.49$^1$ & \\
$^{13}$CH      &  4211.49 & 0.46 & -1.32 &  7.49$^1$ & \\
\ion{O}{i}  &  6300.30 & 0.00 &	-9.94 &  7.45	& \\
\ion{Na}{i} &  5889.95 & 0.00 &  0.12 &  $\le$3.94 & $\le$3.37  \\
\ion{Mg}{i} &  5528.41 & 4.33 &	-0.50 &	 5.36	& 5.40 \\
\noalign{\smallskip}\hline \noalign{\smallskip}
\end{tabular}

{\bf Notes.} \ $^1$ For the $^{12}$C/$^{13}$C = 7 isotope ratio. This table is available in its entirety in a machine-readable form in the online journal. A portion is shown here for guidance regarding its form and content.
\end{table}

\begin{figure}  
 \begin{minipage}{85mm}
\centering
	\includegraphics[width=0.99\textwidth, clip]{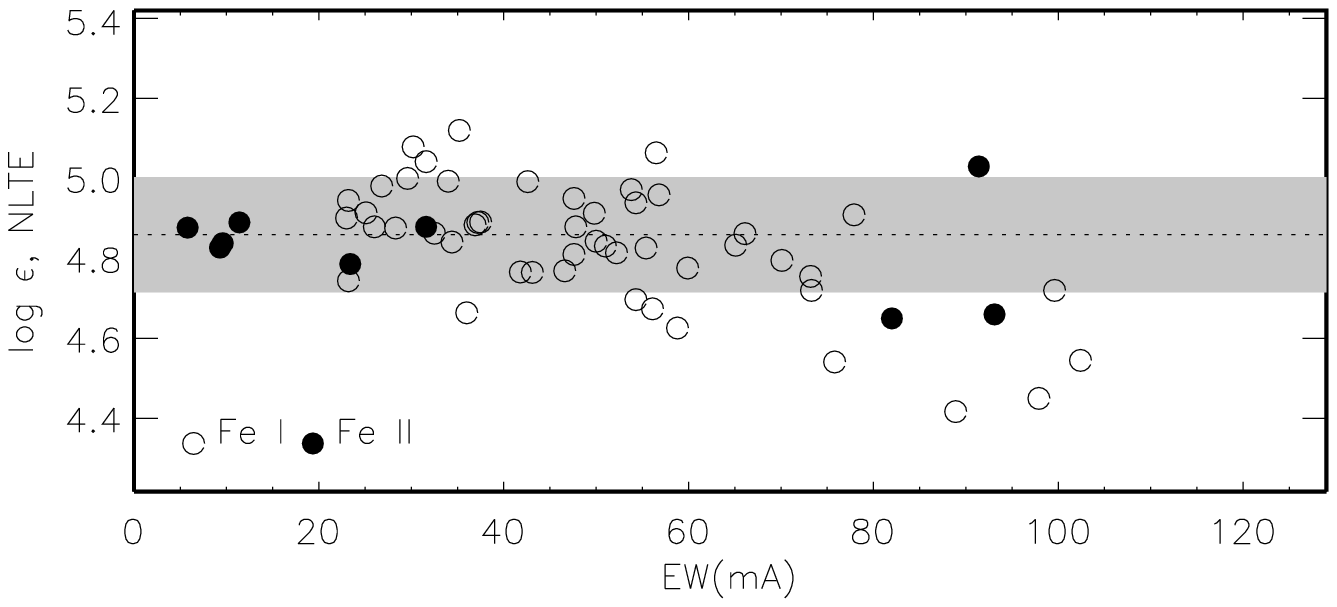}
	\includegraphics[width=0.99\textwidth, clip]{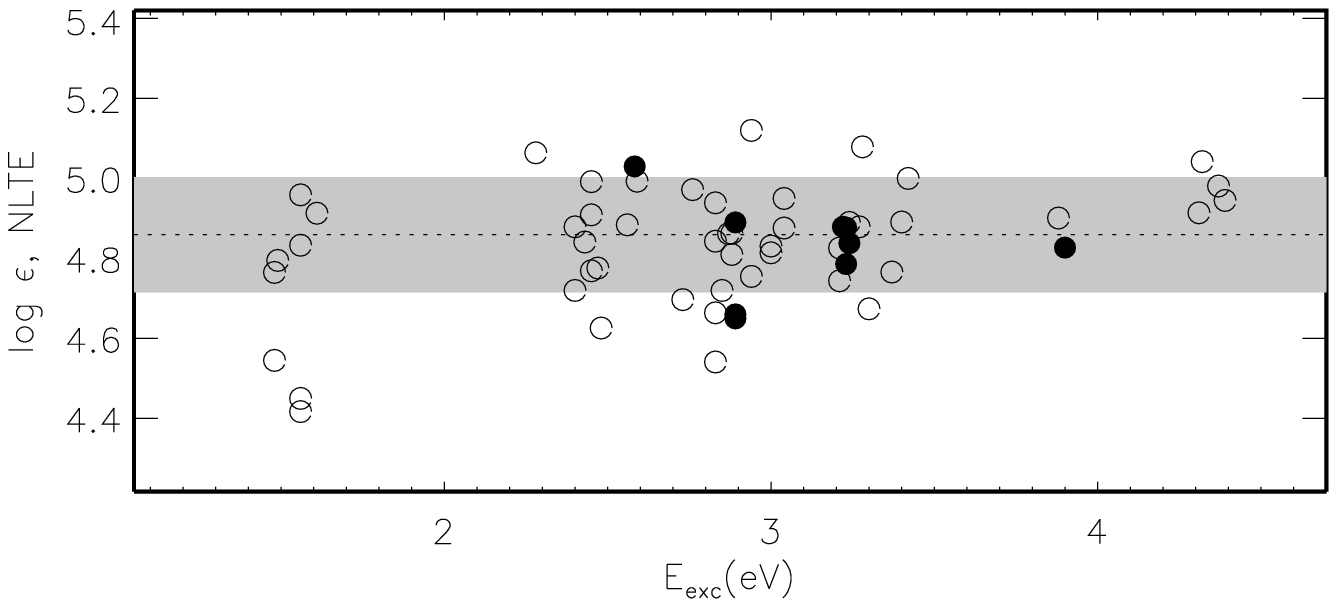}
  \caption{\label{Fig:fe12} NLTE abundances from individual lines of \ion{Fe}{i} (open circles)
    and \ion{Fe}{ii} (filled circles) in Pr184237 as a function of equivalent width EW (top panel) and excitation energy of the lower level (bottom panel). The dotted line indicates the mean \ion{Fe}{i}-based abundance and the shaded grey area its statistical error. }
\end{minipage}
\end{figure}

 We checked the surface gravity $\logg$ = 2.0, which is very close to the FERRE value. The LTE and NLTE abundances are determined from lines of \ion{Fe}{i} and \ion{Fe}{ii} using the model atmosphere with $\Teff$/$\logg$/[Fe/H] = 5100/2.0/$-2.5$ and a microturbulent velocity of $\xi_t$ = 2~\kms, which was calculated with the empirical formula deduced by \citet{dsph_parameters} for metal-poor Galactic halo giants. It provides an accuracy of 0.2~\kms.
The results for individual lines are presented in Table~\ref{Tab:linelist}. The mean \ion{Fe}{i}- and \ion{Fe}{ii}-based NLTE abundances are found to be consistent within 0.02~dex: $\eps{FeI}$ = 4.86$\pm$0.14 (49 lines) and $\eps{FeII}$ = 4.84$\pm$0.12 (9 lines), while a difference in the LTE abundances amounts to $-0.12$~dex. Hereafter, we employ the abundance scale where $\eps{H}$ = 12. The abundance error is calculated as the dispersion in the single line measurements around the mean, $\sigma = \sqrt{\Sigma(\overline{x}-x_i)^2 / (N_l-1)}$. Here, $N_l$ is the number of lines used. Taking into account that two independent methods provide consistent results, we estimate the uncertainty in $\logg$ as 0.1.
Shifts of +0.1/$-0.1$ in $\logg$ result in $\eps{FeI} - \eps{FeII} = -0.02/+0.06$~dex.
Figure~\ref{Fig:fe12} shows no trend of the NLTE abundances with equivalent width, although there are four outliers that reveal a lower abundance compared with the mean, by more than 0.25~dex. The final Fe abundance is computed as the mean of $\varepsilon_{\rm FeI}$ and $\varepsilon_{\rm FeII}$.

\begin{table} 
 \centering
 \caption{\label{Tab:StellarParameter} Determined stellar parameters of Pr184237.}
  \begin{tabular}{lcl}
   \hline\hline
   Parameter & Value & Uncertainty \\
   \hline
   $\Teff$ & 5100 K & $\pm100$~K \\
   $\logg$ & 2.0   & $\pm0.1$\\
   $\mathrm{[Fe/H]}$ & $-2.60$ & $\pm0.14$\\
   $ \xi_t$        & 2.0\,\kms & $ \pm0.2$~\kms \\
   \hline
  \end{tabular}
\end{table}

Despite large reddening for the star observed towards the bulge, we have attempted to derive the effective temperature of our target from the infrared flux method (IRFM) of \citet{2009A&A...497..497G}, and that attempt was successful. We
used the available photometry in the infrared bands from the Two Micron All-Sky Survey \citep[2MASS,][]{1538-3881-131-2-1163}. The 2MASS JHK$_s$ magnitudes
are accurate with J =  13.665$\pm  0.024$, H = 13.114$\pm 0.027$, and K$_s$ = 12.984$\pm 0.030$.
We estimated the Johnson V magnitude (V = 15.942) $\pm 0.012$ converted from the Sloan PanSTARRS
magnitudes g  = 16.370 $\pm 0.003$ and r = 15.550 $\pm 0.001$ using the equation (6) and coefficients from Table~6 in \citet{2012ApJ...750...99T}.
Magnitudes were de-reddened using E(B-V) = 0.38 \citep{green19}. The same reddening value of E(B-V)$_{\rm SFD} = 0.373$ was computed from the dust maps of \citet[][SFD]{1998ApJ...500..525S}.
Finally, E(B-V)$_{\rm SFD}$ was multiplied by a factor of 0.86 following \citet{2011ApJ...737..103S}.
With E(B-V) = 0.321, we obtain $\Teff$(IRFM) = $4999 \pm 102$, $ 5116 \pm 67$, and $5080 \pm 86$~K for the 2MASS
bands J, H, and K$_s$, respectively. The IRFM-based mean temperature of Pr184237,
$\Teff$(IRFM) = $5073 \pm 83$~K, is consistent with that from the H$_\alpha$ wings.
These calculations were performed with $\logg$ = 2 and [Fe/H] = $-2.5$.

We also inferred the effective temperature and surface gravity as in \citet{2023MNRAS.518.4557S}. Briefly, they develop a Monte Carlo method that consists in deriving the effective temperature using the Gaia photometry-temperature relation from \citet{2021A&A...653A..90M} and the surface gravity from the Stephan-Boltzmann equation. This method has been tested to provide compatible stellar parameters with spectroscopic inferences \citep[e.g.][]{2021MNRAS.508.3068L,2023MNRAS.518.4557S}. We find a photometric $\Teff$ = 5190 $\pm$ 100~K and $\logg$ = 2.25 $\pm$ 0.16. The spectroscopic and photometric temperatures agree within 0.63$\sigma$ and the distance-based surface gravity for such a distant object and the spectroscopic one are compatible within 1.32$\sigma$. 

Thus, we adopt $\Teff$ = 5100$\pm100$~K, $\logg$ = 2.0$\pm0.1$, [Fe/H] = $-2.60\pm0.14$, $\xi_t$ = 2$\pm0.2$~\kms\ as the final atmospheric parameters of Pr184237 (Table~\ref{Tab:StellarParameter}).
In order to compute the metallicity, we use the solar Fe abundance $\eps{\odot}$ = 7.45 from \citet{2021SSRv..217...44L}.

\section{Chemical abundances}\label{Sect:abund}

The strongest lithium line, Li\ione\ 6707~\AA, cannot be extracted from noise in the spectrum of Pr184237, suggesting a low Li abundance, as expected for a red giant \citep{1967ApJ...147..624I}. In addition to iron, we are able to determine abundances for 31 chemical elements.
Spectral lines for abundance analysis, together with their atomic data, are selected from the list provided by \citet{HE2327} for an r-process enhanced star, HE~2327-5642, with similar atmospheric parameters. For lines of Sc\ii, Mn\ione, Co\ione, Sr\ii, Ba\ii, La\ii, Eu\ii, and Yb\ii, the HFS and/or IS structure were properly taken into account.

The LTE and NLTE abundances obtained from individual lines are presented in Table~\ref{Tab:linelist}. Table~\ref{Tab:abund} lists the elemental average abundances together with their stochastic errors. Stochastic errors are caused by random uncertainties in the continuum placement, line profile fitting, and $gf$-values. They are represented by the dispersion in the measurements $\sigma$ when $N_l > 1$. When only a single line or a molecular band is measured, the abundance error resembles the fit uncertainty provided by MPFIT \citep{2009ASPC..411..251M} and computed as $PERROR \cdot \sqrt{\chi^2/N_{DOF}}$, where $N_{DOF}$ is the number of fitting degrees of freedom and $PERROR$ is a formal parameter uncertainty, computed from the covariance matrix.
In order to compute the abundance ratios relative to iron, [X/Fe], we used the present Solar system abundances recommended by \citet{2021SSRv..217...44L} and [Fe/H] = $-2.60$.
For Na, Mg, Al, Si, Ca, Sc, Ti, and Zn, the [X/Fe] ratios are based on the NLTE abundances. For Cr, Mn, Co, and Ni, Table~\ref{Tab:abund} indicates the LTE abundance ratio [X/Fe\ione], which is free, in part, from the departures from LTE. Since the NLTE abundances were determined for 4 of 15 heavy elements, the [X/Fe] ratios for all the heavy elements in Table~\ref{Tab:abund} and the heavy element abundance pattern in our further analysis (Sections~\ref{sect:saga} and \ref{Sect:origin}) are based on the LTE abundances, for consistency. In Sect~\ref{sect:uncert}, we discuss the influence of using the NLTE abundances on our analysis of the nucleosynthesis scenarios.

\begin{table} 
 \caption{\label{Tab:abund} Elemental abundances of Pr184237.}
 \centering
 \begin{tabular}{lcrrrr}
\hline\hline \noalign{\smallskip}
Species  & $\eps{\odot}$ & \multicolumn{2}{c}{$\eps{}$} & $N_l$ & \multicolumn{1}{c}{[X/Fe]} \\
\cline{3-4} \noalign{\smallskip}
         &               & \multicolumn{1}{c}{LTE} & \multicolumn{1}{c}{NLTE} & & \\
\noalign{\smallskip} \hline \noalign{\smallskip}
C(C$_2$)& 8.47     &   7.64(0.03) &            & 1  &   1.77   \\
N(CN)   & 7.85     &   6.54(0.10) &            & 1  &   1.29   \\
O\ione  & 8.73     &   7.45(0.03) &            & 1  &   1.32   \\
Na\ione & 6.27     & $<$ 3.94   & $<$ 3.37   & 2  & $< -0.30^{\rm N}$ \\
Mg\ione & 7.52     &   5.36(0.02) & 5.40(0.02) & 1  &   0.48$^{\rm N}$   \\
Al\ione & 6.42     &   3.31(0.15) & 3.76(0.15) & 1  & --0.06$^{\rm N}$   \\
Si\ione & 7.51     &   5.30(0.10) & 5.27(0.10) & 1  &   0.36$^{\rm N}$   \\
Ca\ione & 6.27     &   4.04(0.11) & 4.14(0.11) & 7  &   0.47$^{\rm N}$   \\
Sc\ii   & 3.04     &   0.90(0.05) & 0.90(0.05) & 1  &   0.46$^{\rm N}$   \\
Ti\ii   & 4.90     &   2.72(0.12) & 2.72(0.12) & 9  &   0.42$^{\rm N}$   \\
V\ii    & 3.95     &   1.28(0.05) &            & 2  & --0.08   \\
Cr\ione & 5.63     &   2.81(0.06) &            & 3  & --0.09$^{\rm I}$   \\
Mn\ione & 5.47     &   2.41(0.10) &            & 1  & --0.33$^{\rm I}$   \\
Fe\ione & 7.45     &   4.72(0.13) & 4.86(0.14) & 49 &   0.01$^{\rm N}$   \\
Fe\ii   & 7.45     &   4.84(0.12) & 4.84(0.12) & 9  & --0.01$^{\rm N}$   \\
Co\ione & 4.86     &   2.44(0.05) &            & 3  &   0.31$^{\rm I}$   \\
Ni\ione & 6.20     &   3.40(0.18) &            & 5  & --0.07$^{\rm I}$   \\
Zn\ione & 4.61     &   2.05(0.10) & 2.22(0.10) & 1  &   0.21$^{\rm N}$   \\
Sr\ii   & 2.88     &   0.27(0.05) & 0.22(0.05) & 2  & --0.02   \\
Y\ii    & 2.15     & --0.24(0.20) &            & 8  &   0.21   \\
Zr\ii   & 2.55     &   0.67(0.22) &            & 4  &   0.72   \\
Ba\ii   & 2.17     &   0.74(0.08) & 0.40(0.10) & 3  &   1.17   \\
La\ii   & 1.17     & --0.33(0.13) &            & 7  &   1.10   \\
Ce\ii   & 1.58     &   0.09(0.19) &            & 5  &   1.11   \\
Nd\ii   & 1.45     & --0.08(0.14) &            & 10 &   1.07   \\
Sm\ii   & 0.94     & --0.55(0.13) &            & 2  &   1.11   \\
Eu\ii   & 0.51     & --1.24(0.06) & --1.09(0.06) & 4 & 0.85   \\
Gd\ii   & 1.05     & --0.77(0.08) &            & 2  &  0.78   \\
Dy\ii   & 1.12     & --0.47(0.09) &            & 5  &  1.01   \\
Er\ii   & 0.92     & --0.67(0.07) &            & 3  &  1.01   \\
Tm\ii   & 0.11     & --1.00(0.05) &            & 1  &  1.49   \\
Yb\ii   & 0.91     & --0.66(0.32) &            & 1  &  1.03   \\
Pb\ione & 2.03     &   1.55(0.16) & 2.21(0.16) & 2  &  2.12   \\
\noalign{\smallskip}\hline \noalign{\smallskip}
\end{tabular}
{\bf Notes.} The numbers in parentheses are the abundance errors. $^{\rm N}$ Based on the NLTE abundance. $^{\rm I}$ LTE abundance ratio [X/Fe\ione].
\end{table}

We examined the systematic uncertainties in the derived elemental abundances of Pr184237, which are linked to the adopted stellar parameters. These were estimated by varying $\Teff$ by 100~K, $\logg$ by 0.1, and $\xi_t$ by $-0.2$~\kms\ in the model atmosphere.
Table~\ref{Tab:uncertainty} summarizes the various sources of uncertainties. The total impact of varying $\Teff$, $\logg$, and $\xi_t$ is computed as the quadratic sum of Cols.~2-4 and denoted $\Delta(T,g,\xi_t)$.
 Column~6 lists the stochastic errors $\sigma_{\rm obs}$, as given in Table~\ref{Tab:abund}.
The total uncertainty $\sigma_{\rm tot}$ in the absolute abundance of each element is computed by the quadratic sum of the stochastic and systematic (Col.~5) errors.

\begin{table} 
 \caption{\label{Tab:uncertainty} Error budget for elemental abundances in Pr184237.}
 \centering
 \begin{tabular}{lcrrccc}
\hline\hline \noalign{\smallskip}
El. & $\Delta T$ & \multicolumn{1}{c}{$\Delta \logg$} & \multicolumn{1}{c}{$\Delta \xi_t$} & $\Delta$ & $\sigma_{\rm obs}$ & $\sigma_{\rm tot}$ \\
    & 100 K & \multicolumn{1}{c}{0.1} & --0.2 \kms & ($T,g,\xi_t$) &  &  \\
(1) & (2) & \multicolumn{1}{c}{(3)} & \multicolumn{1}{c}{(4)} & (5) & (6) & (7) \\
\noalign{\smallskip} \hline \noalign{\smallskip}
C$_2$   &  0.14 & -0.02 &  $<$0.01 &  0.14 &  0.03 &  0.14 \\
CN      &  0.17 & -0.02 &  $<$0.01 &  0.17 &  0.10 &  0.20 \\
O\ione  &  0.07 &  0.03 &  $<$0.01 &  0.08 &  0.03 &  0.08 \\
Mg\ione &  0.06 &  $<$0.01 &  0.03 &  0.07 &  0.02 &  0.07 \\
Al\ione &  0.11 &  $<$0.01 &  0.10 &  0.15 &  0.15 &  0.21 \\
Si\ione &  0.11 &  $<$0.01 &  0.05 &  0.12 &  0.10 &  0.16 \\
Ca\ione &  0.07 &  $<$0.01 &  0.03 &  0.08 &  0.11 &  0.13 \\
Sc\ii &  0.05 &  0.04 &  0.02 &  0.07 &  0.05 &  0.08 \\
Ti\ii &  0.05 &  0.04 &  0.02 &  0.07 &  0.12 &  0.14 \\
V\ii  &  0.05 &  0.04 &  0.01 &  0.06 &  0.05 &  0.08 \\
Cr\ione &  0.11 &  $<$0.01 &  0.02 &  0.11 &  0.06 &  0.13 \\
Mn\ione &  0.11 &  $<$0.01 &  0.05 &  0.12 &  0.10 &  0.16 \\
Fe\ione &  0.11 &  $<$0.01 &  0.05 &  0.12 &  0.13 &  0.18 \\
Fe\ii   &  0.02 &  0.04 &  0.04 &  0.06 &  0.12 &  0.13 \\
Co\ione &  0.11 &  $<$0.01 &  $<$0.01 &  0.11 &  0.05 &  0.12 \\
Ni\ione &  0.11 &  $<$0.01 &  0.04 &  0.12 &  0.18 &  0.21 \\
Zn\ione &  0.07 &  0.01 &  $<$0.01 &  0.07 &  0.10 &  0.12 \\
Sr\ii &  0.09 &  0.01 &  0.11 &  0.14 &  0.05 &  0.15 \\
Y\ii  &  0.06 &  0.03 &  0.03 &  0.07 &  0.20 &  0.21 \\
Zr\ii &  0.07 &  0.03 &  0.03 &  0.08 &  0.22 &  0.23 \\
Ba\ii &  0.09 &  0.03 &  0.14 &  0.17 &  0.08 &  0.19 \\
La\ii &  0.09 &  0.03 &  $<$0.01 &  0.09 &  0.13 &  0.16 \\
Ce\ii &  0.09 &  0.03 &  $<$0.01 &  0.09 &  0.19 &  0.21 \\
Nd\ii &  0.09 &  0.03 &  0.01 &  0.10 &  0.14 &  0.17 \\
Sm\ii &  0.09 &  0.03 &  $<$0.01 &  0.09 &  0.13 &  0.16 \\
Eu\ii &  0.09 &  0.03 &  $<$0.01 &  0.09 &  0.06 &  0.11 \\
Gd\ii &  0.09 &  0.03 &  $<$0.01 &  0.09 &  0.08 &  0.12 \\
Dy\ii &  0.09 &  0.03 &  0.01 &  0.10 &  0.09 &  0.13 \\
Er\ii &  0.09 &  0.03 &  $<$0.01 &  0.09 &  0.07 &  0.12 \\
Tm\ii &  0.09 &  0.03 &  $<$0.01 &  0.09 &  0.05 &  0.11 \\
Yb\ii &  0.06 &  0.03 &  0.09 &  0.11 &  0.32 &  0.34 \\
Pb\ione &  0.11 &  $<$0.01 &  0.02 &  0.11 &  0.16 &  0.20 \\
\noalign{\smallskip} \hline
\end{tabular}
\end{table}

The results for different groups of elements are described below.

\subsection{Light elements C, N, O and isotope ratio $^{12}$C/$^{13}$C}\label{sect:cno}

\begin{figure}  
 \begin{minipage}{85mm}
\centering
	\includegraphics[width=0.99\textwidth, clip]{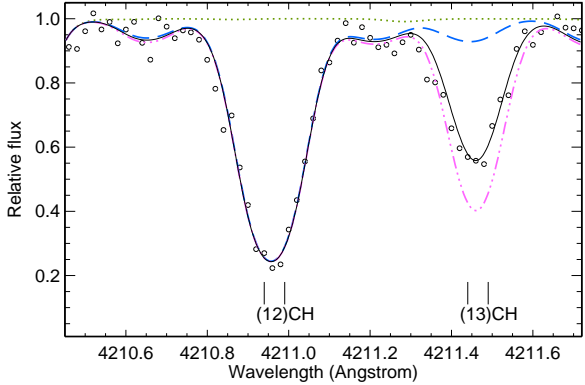}
  \caption{\label{Fig:ch12_13} Molecular $^{12}$CH and $^{13}$CH (denoted as (12)CH and (13)CH) lines in spectrum of Pr184237 (open circles). The best fit (continuous curve) to the observations was achieved for the $^{12}$C/$^{13}$C = 7 isotopic ratio and $\eps{C}$ = 7.49. The blue dashed and magenda three-dot-dashed curves correspond to  $^{12}$C/$^{13}$C = 95 (the solar ratio) and 4, respectively. In two latter cases, the total carbon abundances are $\eps{C}$ = 7.44 and 7.54. }
\end{minipage}
\end{figure}

The FERRE analysis discovered a strong enhancement of Pr184237 in carbon.
In this study, we use a pronounced C$_2$ Swan band (5150 -- 5170~\AA) and determine the C abundance with an uncertainty of 0.03~dex.
We confirm the FERRE conclusion although obtain the lower [C/Fe] = 1.77 due to adopting the lower effective temperature.

New results are measurements for the N and O abundances. With fixed C abundance, the N abundance was determined from the CN band in the 4214-4216\,\AA\ region. For oxygen, we use the [O\ione] 6300~\AA\ forbidden line, which is clearly detected in the spectrum of Pr184237. The star is found to be strongly enhanced in N and O, with [N/Fe] = 1.29 and [O/Fe] = 1.32.

 Owing to the high carbon abundance, we derive the $^{12}$C/$^{13}$C isotope abundance ratio from pairs of $^{12}$CH 4210.9~\AA\ and $^{13}$CH 4211.4~\AA\ molecular lines (Fig.~\ref{Fig:ch12_13}). The calculations are performed with the solar mixture of C isotopes, $^{12}$C/$^{13}$C = 95 \citep{2021SSRv..217...44L}, and with $^{12}$C/$^{13}$C = 9, 7, and 4. The best fit to the observed spectrum was achieved for $^{12}$C/$^{13}$C = 7. A variation in $^{12}$C/$^{13}$C only influences a little the theoretical profile of $^{12}$CH 4210.9~\AA, such that the C abundance obtained from fitting this line increases from $\eps{C}$ = 7.44 to 7.54, when $^{12}$C/$^{13}$C decreases from 95 ($^{12}$C fraction of 98.965\%) to 4 ($^{12}$C fraction of 80\%).

A supersolar fraction of the $^{13}$C isotope can be a signature of AGB nucleosynthesis \citep[e.g.,][]{karakas10,2012ApJ...747....2L,ventura21}.
Here, the donor AGB star would likely have been of intermediate-mass ($M \gtrsim 2 M_{\odot}$), such that it experienced proton capture nucleosynthesis at the convective envelope, leading to a low $^{12}$C/$^{13}$C ratio and a large overabundance of N. Such stars are important contributors to  $^{13}$C and N in the early Galaxy \citep{2020ApJ...900..179K}. Low-metallicity AGB stars can also produce large overabundances of O, which are dredged to the surface alongside C and heavy elements \citep{karakas10,2012ApJ...747....2L}. 
%
With $\Teff$ = 5100~K and $\logg$ = 2.0, Pr184237 is far from the AGB evolutionary phase, and the self-pollution hypothesis should be rejected. 
 Variations in the radial velocity suggest binarity of the system, where Pr184237 is a less massive secondary component. The primary evolved to the AGB phase and via the mass transfer provided Pr184237 with the nuclear-processed material. At present, the primary is, probably, a white dwarf.



\subsection{Elements Na to Zn}

Lines of the four $\alpha$-process elements, namely, Mg\ione, Si\ione, Ca\ione, and Ti\ii\ are measured in the spectrum of Pr184237. The star reveals an $\alpha$-enhancement with tight measurements of [X/Fe] = 0.48, 0.36, 0.47, and 0.42, respectively. These numbers correspond to the NLTE calculations.

Pronounced NLTE effects are found for the resonance lines of Na\ione\ and Al\ione, such that NLTE leads to the lower abundance of Na compared with the LTE case, by 0.57~dex, and, in contrast, to a higher abundance of Al, by 0.45~dex.
Red wings of the Na\ione\ 5889 and 5895\,\AA\ lines in spectrum of Pr184237 are affected by interstellar absorption lines, and we can only evaluate an upper limit for the element abundance: [Na/Fe] $\le -0.3$ in the NLTE calculations. It is important to note that Na is certainly depleted relative to Fe and does not follow the enhanced carbon.
Aluminum does not reveal a notable deviation from the scaled solar abundance, with [Al/Fe] = $-0.06$ in the NLTE calculations.

Abundances of Sc and V were determined from lines of their majority species, Sc\ii\ and V\ii, which are expected to be weakly affected by the departures from LTE. Indeed, the NLTE calculations resulted in $\Delta_{\rm NLTE} < 0.01$~dex for the Sc\ii\ 4246~\AA\ line observed in Pr184237.
Scandium appears to be enhanced relative to Fe to the same extent, as are the $\alpha$-process elements.
Vanadium follows iron, with [V/Fe] = $-0.08$. 

Abundances of Cr, Mn, Co, and Ni were derived from lines of their minority species, which are subject to the ultraviolet overionization, resulting in weakened lines and positive NLTE abundance corrections.
For Cr\ione\ and Mn\ione, the NLTE corrections are available in the database NLTE\_MPIA\footnote{\tt http://nlte.mpia.de/} \citep{NLTE_MPIA}. They were computed based on the methods of \citet{2010A&A...522A...9B} and \citet{2019A&A...631A..80B}, respectively.
For the atmospheric parameters of Pr184237, we extracted $\Delta_{\rm NLTE}$ = 0.56~dex, 0.58~dex, 0.48~dex, and 0.26~dex for Cr\ione\ 4254, 4274, 5409~\AA\ and Mn\ione\ 4033~\AA, respectively, and obtained [Cr/Fe](NLTE) = 0.33 and [Mn/Fe](NLTE) = $-0.20$. Thus, manganese in Pr184237 is deficient relative to Fe, independent of whether we rely on NLTE or LTE. Overabundance of Cr relative to Fe is unlikely. When taking the Cr\ione - and Fe\ione -based LTE abundances, we obtained a close-to-solar ratio of [Cr\ione /Fe\ione] = $-0.09$.

For the Co\ione\ lines, the NLTE corrections are not available in \citet{2010MNRAS.401.1334B} and, currently, not accessible at the above cited website. However, they are certainly positive. Therefore, cobalt is enhanced in Pr184237, at the level of [Co/Fe] $>$ 0.2.

When taking the LTE abundances from lines of Ni\ione\ and Fe\ione, we obtained [Ni\ione /Fe\ione] = $-0.07$, suggesting that the Ni abundance in Pr184237 is close to the scaled solar one.

Zinc is slightly enhanced, with [Zn/Fe](NLTE) = 0.21.


\subsection{Neutron-capture elements Sr to Pb}\label{sect:heavy}

\begin{figure}  
 \begin{minipage}{85mm}
\centering
	\includegraphics[width=0.99\textwidth, clip]{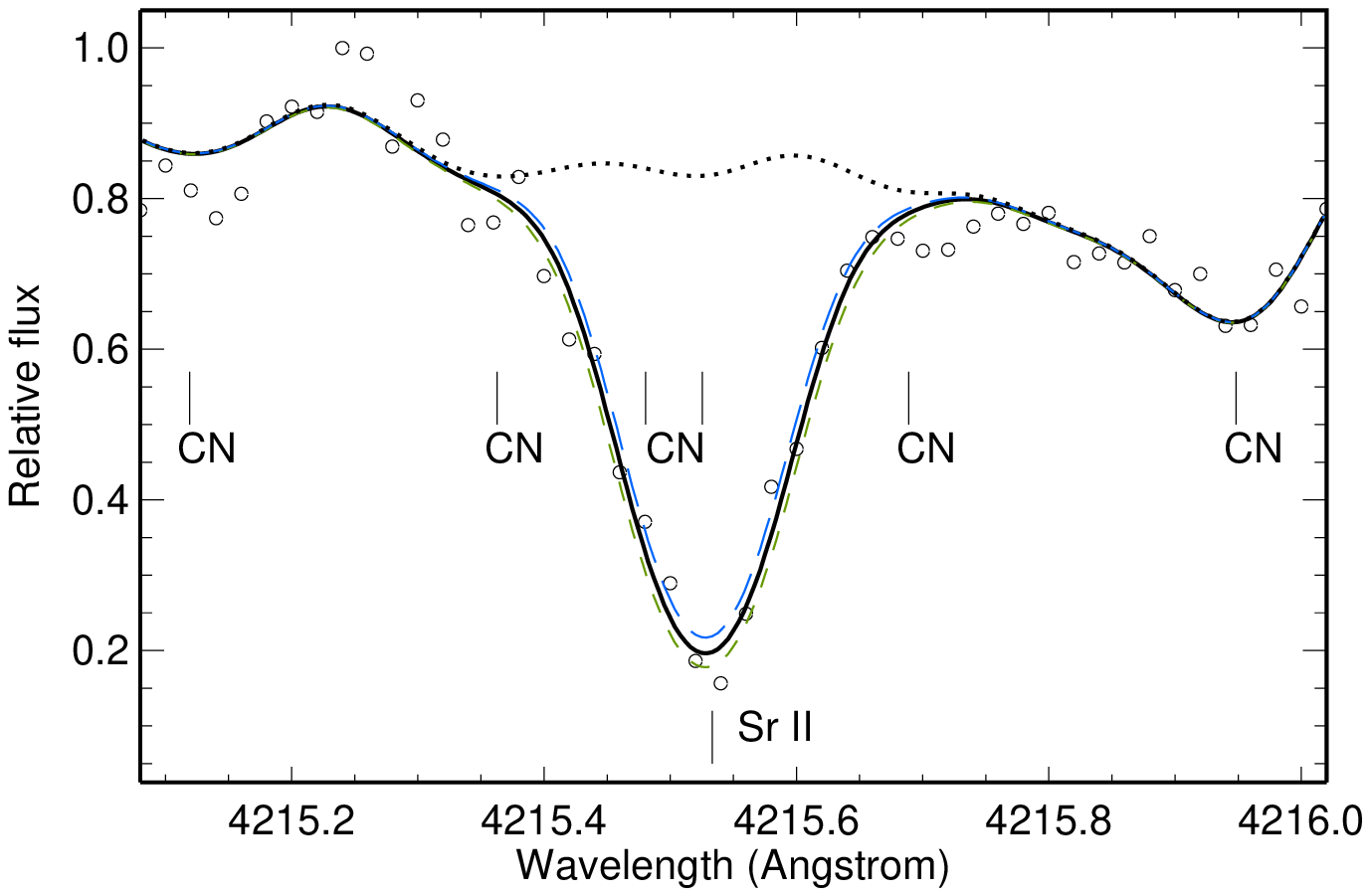}
	\includegraphics[width=0.99\textwidth, clip]{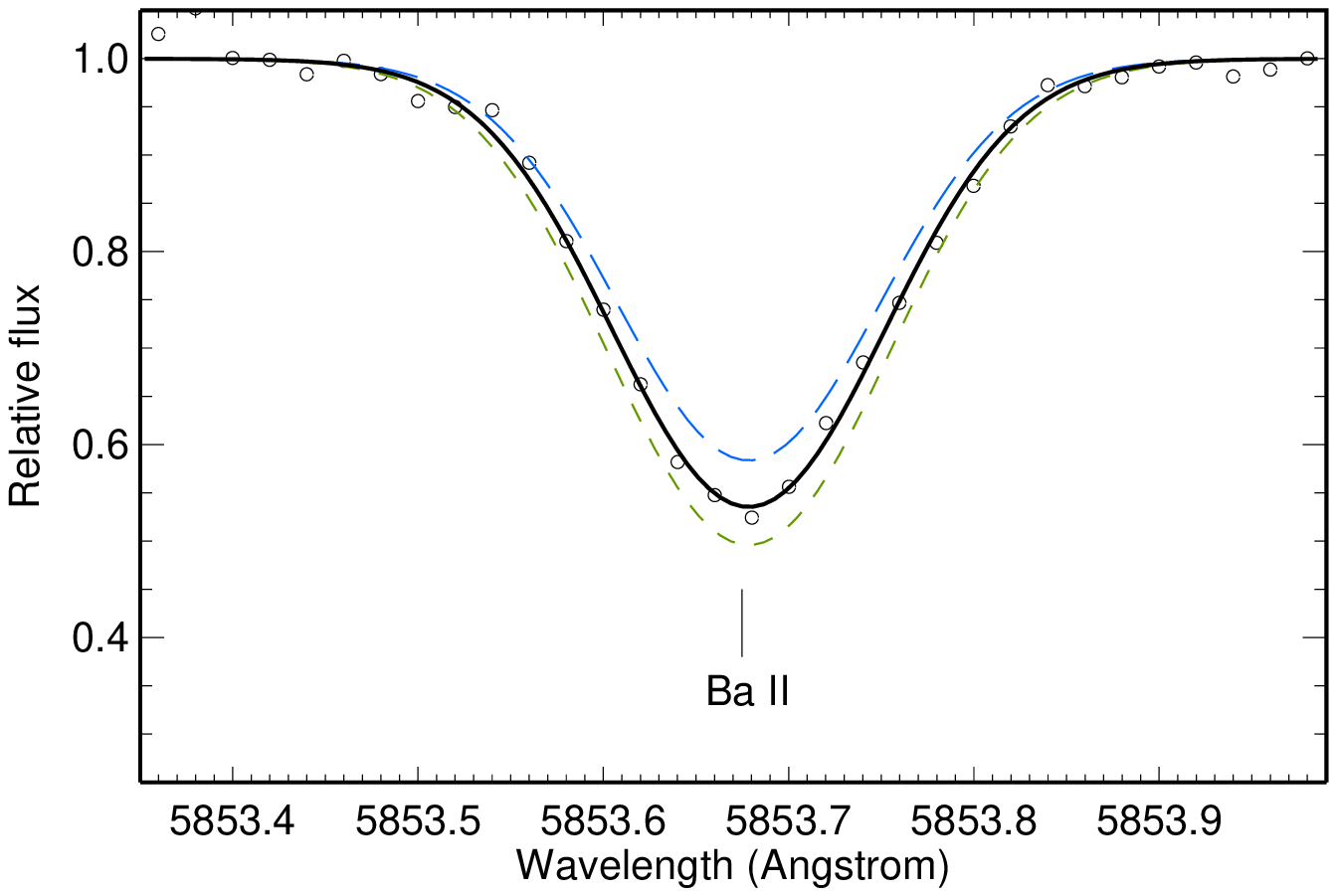}
	\includegraphics[width=0.99\textwidth, clip]{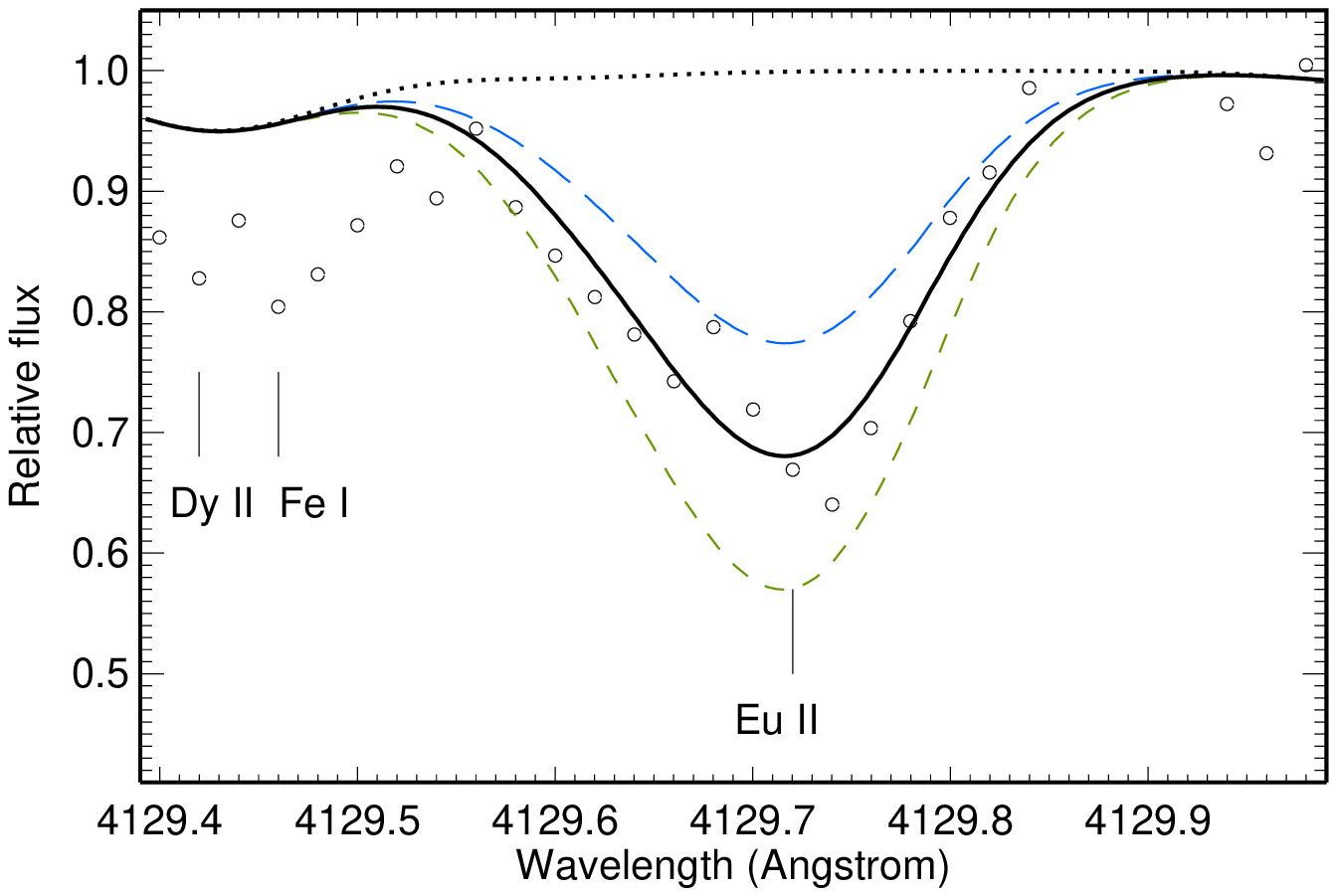}
  \caption{\label{Fig:sr_eu} Best fits (continuous curve) to the Sr\ii\ 4215\,\AA\ (top panel), Ba\ii\ 5853\,\AA\ (middle panel), and Eu\ii\ 4129\,\AA\ (bottom panel) lines in Pr184237 (open circles). The derived abundances are indicated in Table~\ref{Tab:linelist}. The dotted curves in the top and bottom panels show the synthetic spectra with no Sr and Eu in the atmosphere. The blue long-dashed and green short-dashed curves show the effect of a 0.2~dex variation in the abundance on the synthetic spectrum.}
\end{minipage}
\end{figure}

We determine abundance for 15 elements beyond Sr (hereafter, heavy elements). Seven of them, namely, Sr, Y, Zr, Ba, La, Ce, and Pb, are commonly referred to as s-process elements because of the dominant contribution of the s-process to their solar abundances, of greater than 60\%\ according to \citet{2014ApJ...787...10B}. Five measured elements, namely, Eu, Gd, Dy, Er, and Tm, are presumably of r-process origin, with more than 80\%\ contribution of the r-process to their solar abundances. The remaining elements Nd, Sm, and Yb are intermediate cases, with approximately equal contributions of the r- and s-process to their solar abundances.

For Sr to Er, their abundances are derived from two to ten lines. The best fits to the representative lines are shown in Fig.~\ref{Fig:sr_eu}.
Ytterbium is observed in the only line, Yb\ii\ 3694~\AA, and the only unblended line was found for Tm\ii.

\begin{figure}  
 \begin{minipage}{85mm}
\centering
	\includegraphics[width=0.99\textwidth, clip]{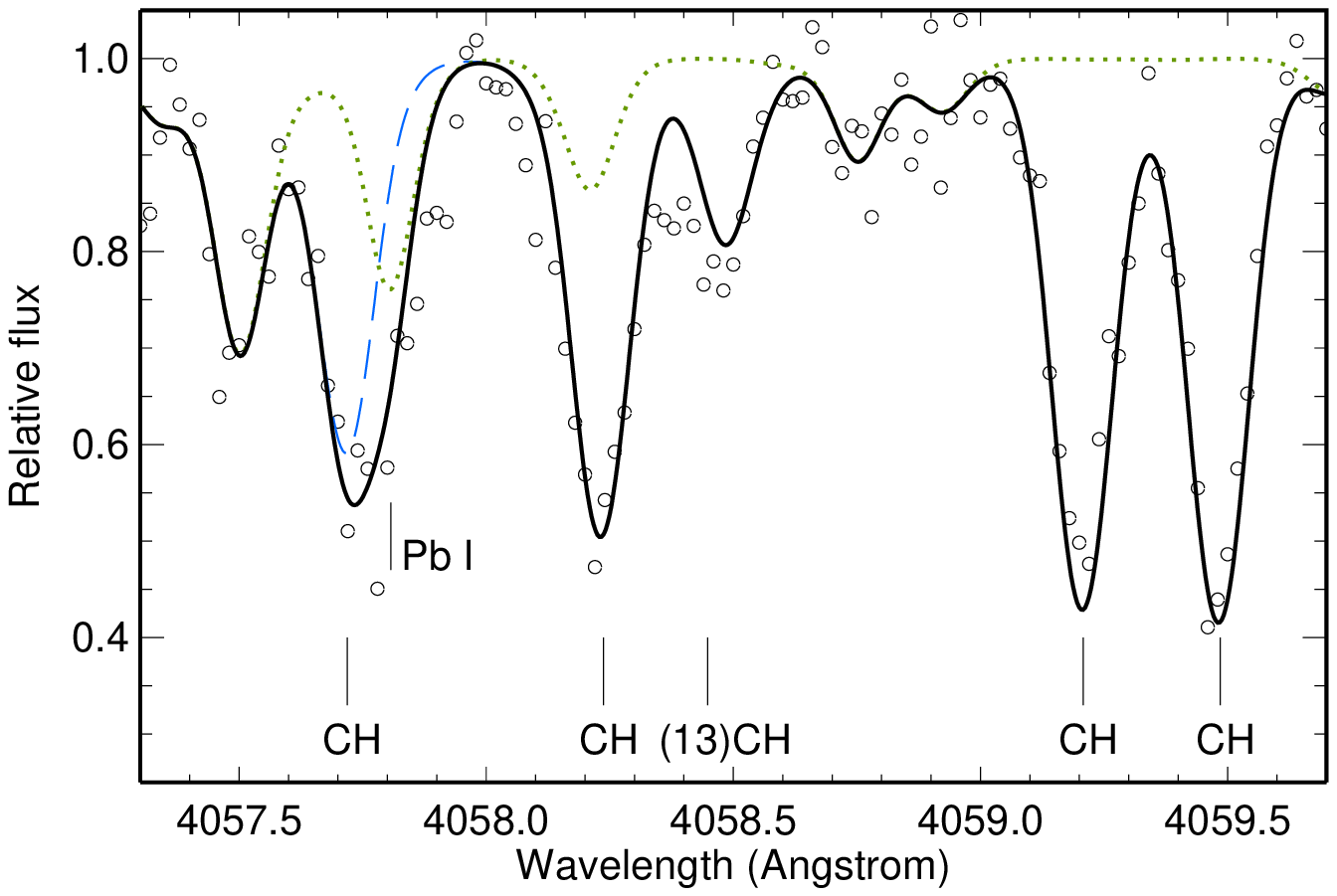}
	\includegraphics[width=0.99\textwidth, clip]{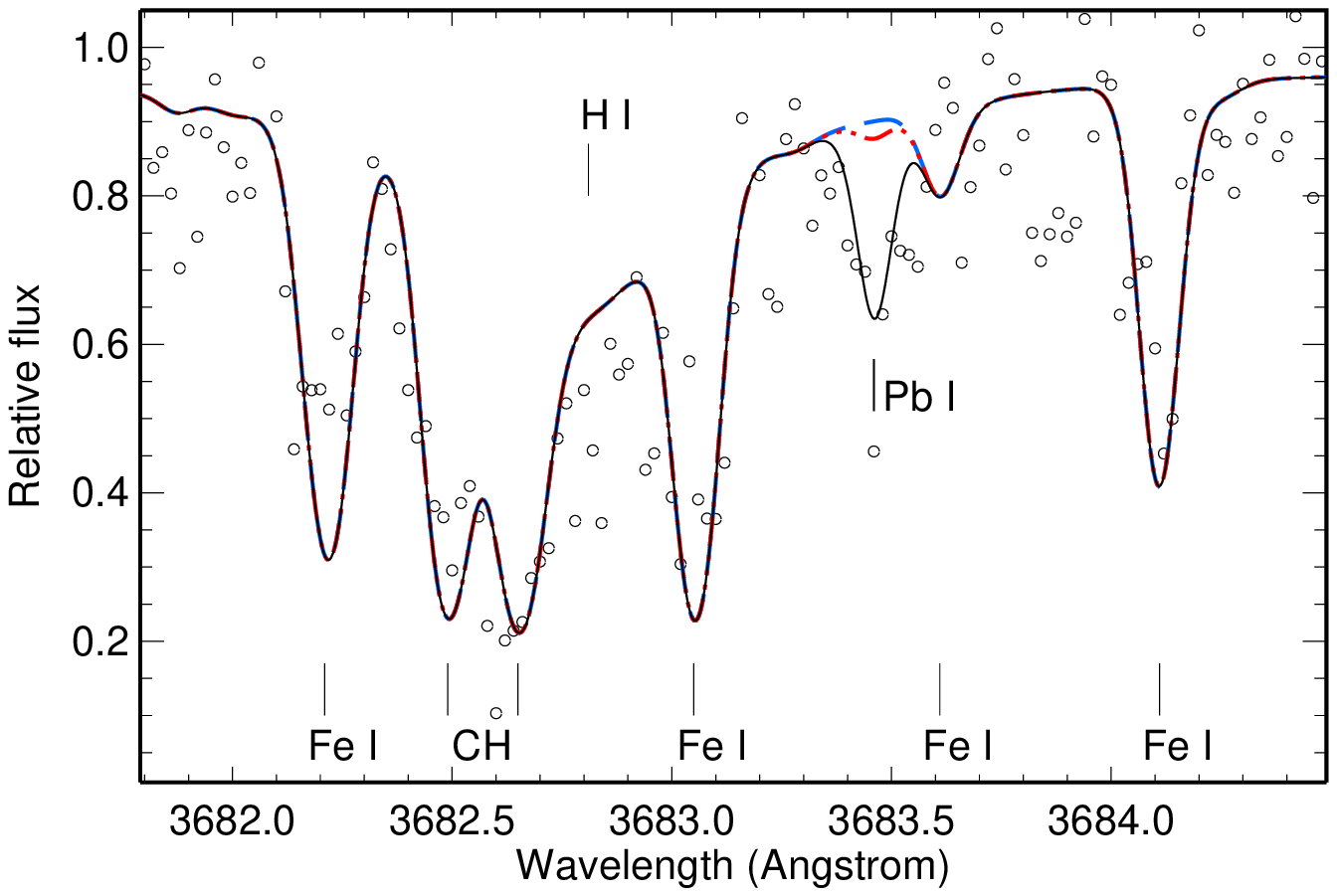}
  \caption{\label{Fig:pb} Best fits (continuous curve) to the Pb\ione\ 4057\,\AA\ (top panel) and 3683\,\AA\ (bottom panel) lines in Pr184237 (open circles). The derived abundances are indicated in Table~\ref{Tab:linelist}. The blue dashed curves show the synthetic spectra computed with [Pb/Fe] = 0. The red dash-dotted curve in the bottom panel corresponds to [Pb/Fe] = 1.0. The green dotted curve in the top panel shows the synthetic spectrum computed without carbon in the atmosphere. }
\end{minipage}
\end{figure}

A challenge was to determine the Pb abundance from the Pb\ione\ 4057\,\AA\ line, which is heavily blended by the molecular CH line. As shown in Fig.~\ref{Fig:pb} (top panel), there are three unblended or weakly blended lines of $^{12}$CH in the 4058.0-4059.7~\AA\ range. They all are well fitted with $\eps{C}$ = 7.58, which agrees within 0.06~dex with the C abundance derived from the C$_2$ Swan band. We note that the $^{13}$CH 4058.448~\AA\ line is reproduced reasonably well using $^{12}$C/$^{13}$C = 7, as derived in this study. It is evident that neither the position, nor the total absorption of the 4057~\AA\ blend can be reproduced in the case of [Pb/Fe] = 0 ($\eps{Pb} = -0.57$). In the LTE calculations, the best fit to the 4057~\AA\ blend is achieved with $\eps{Pb}$ = 1.55. The obtained Pb abundance is supported by another line of Pb\ione, at 3683\,\AA. This line is located in a rather crowded region (Fig.~\ref{Fig:pb} bottom panel), which is nevertheless well fitted using the derived atmospheric parameters (important for H\ione\ 3682.8~\AA !) and abundances of C and Fe. For Pb\ione\ 3683\,\AA, we show the synthetic profiles computed with [Pb/Fe] = 0, 1, and 2.12. It is evident that the first two profiles are too shallow compared with the observed one. Only Pb\ione\ 4057\,\AA\ was used to represent in Table~\ref{Tab:uncertainty} the systematic uncertainties in the Pb abundance. When varying $\Teff$ or $\logg$ or $\xi_t$, we first evaluated the C abundance from the CH 4058.2, 4059.2, 4059.5~\AA\ lines, and then the Pb abundance was obtained from fitting the 4057.7~\AA\ blend.

We obtain that elements of the second s-process peak (Ba, La, Ce) and elements of the intermediate group (Nd, Sm, Yb) are strongly enhanced in Pr184237, with [X/Fe] $> 1$.
An even higher excess was found for Pb, with [Pb/Fe] = 2.12 and 2.78 in the LTE and NLTE calculations, respectively. 

Pr184237 is the first star in the inner Galaxy for which the five r-process elements are measured, and they all reveal an enhancement, with the average [r/Fe] = 1.03$\pm$0.28.

\begin{figure}  
 \begin{minipage}{85mm}
\centering
	\includegraphics[width=0.99\textwidth, clip]{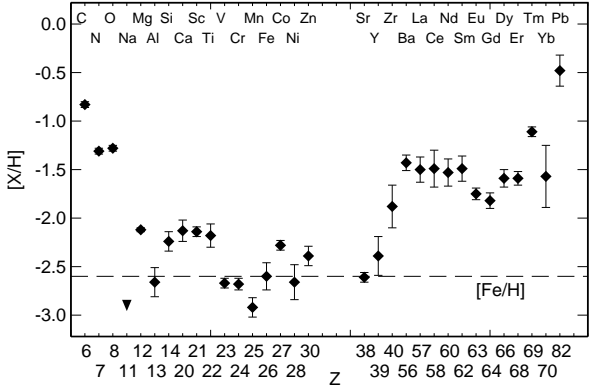}
  \caption{\label{Fig:pattern} The element abundance pattern of Pr184237. An upper limit for the Na abundance is indicated by downward-facing triangle. The dashed line indicates the star’s iron abundance [Fe/H] = $-2.60$. Abundances of Cr, Mn, Co, and Ni are increased by 0.14~dex, see text for more details.}
\end{minipage}
\end{figure}

The obtained element abundance pattern of Pr184237 is presented in Fig.~\ref{Fig:pattern}. For illustration purpose, the LTE abundances of Cr, Mn, Co, and Ni were increased by 0.14~dex, which is the difference between the NLTE and LTE abundances for Fe\ione. 
By doing so, we assumed similar NLTE effects for neutral species of these five elements and, in the first approximation, accounted for the NLTE effects for Cr\ione, Mn\ione, Co\ione, and Ni\ione.
 For all elements beyond Sr, the LTE abundances are used for consistency.

\subsection{Comparisons with the VMP stellar samples}\label{sect:saga}

For abundance comparisons, we use the VMP stellar samples observed towards the bulge from high-resolution spectroscopic studies of \citet{2016MNRAS.460..884H}, \citet{2016A&A...587A.124K}, and \citet{2023MNRAS.518.4557S}. 

Both \citet{2016A&A...587A.124K} and \citet{2023MNRAS.518.4557S} conclude that the majority of their carbon-normal stars exhibit abundance patterns typical for the Galactic halo stars, although the [Na/Mg] ratios in \citet{2023MNRAS.518.4557S} reveal a substantial star-to-star scatter. In contrast, \citet{2016MNRAS.460..884H} find that the chemistry of their sample stars deviates from that for halo stars of the same metallicity. However, we see in their Figures~11-13 that most of the elemental ratios lie well on the halo abundance trends. The star-to-star scatter is, indeed, observed for [Mg/Fe].
In the Na–Zn range, our CEMP star Pr184237 does not look exceptional among the non-CEMP stars.

Compared with the three carbon and s-process enhanced
stars from \citet{2016A&A...587A.124K} and \citet{2023MNRAS.518.4557S}, Pr184237 has the lowest [Ba/Fe] = 1.17, however, this is the first star discovered in the inner Galaxy that is strongly enhanced in both s- and r-process elements. Not only Eu, but also Gd, Dy, Er, and Tm were measured in this star. With [Ba/Eu] = $0.32$, Pr184237 likely belongs to the CEMP-r/s subclass.

Pr184237 is the only CEMP star in the inner Galaxy, for which the isotope abundance ratio $^{12}$C/$^{13}$C and the Pb abundance are derived. Earlier, the Pb abundance, [Pb/Fe] = 1.50, was measured for a more metal-rich and non-CEMP star with [Fe/H] = $-1.5$ and [C/Fe] = 0.4 \citep{2019A&A...622A.159K}.

\begin{figure*}  
 \begin{minipage}{170mm}
\centering
	\includegraphics[width=0.33\textwidth, clip]{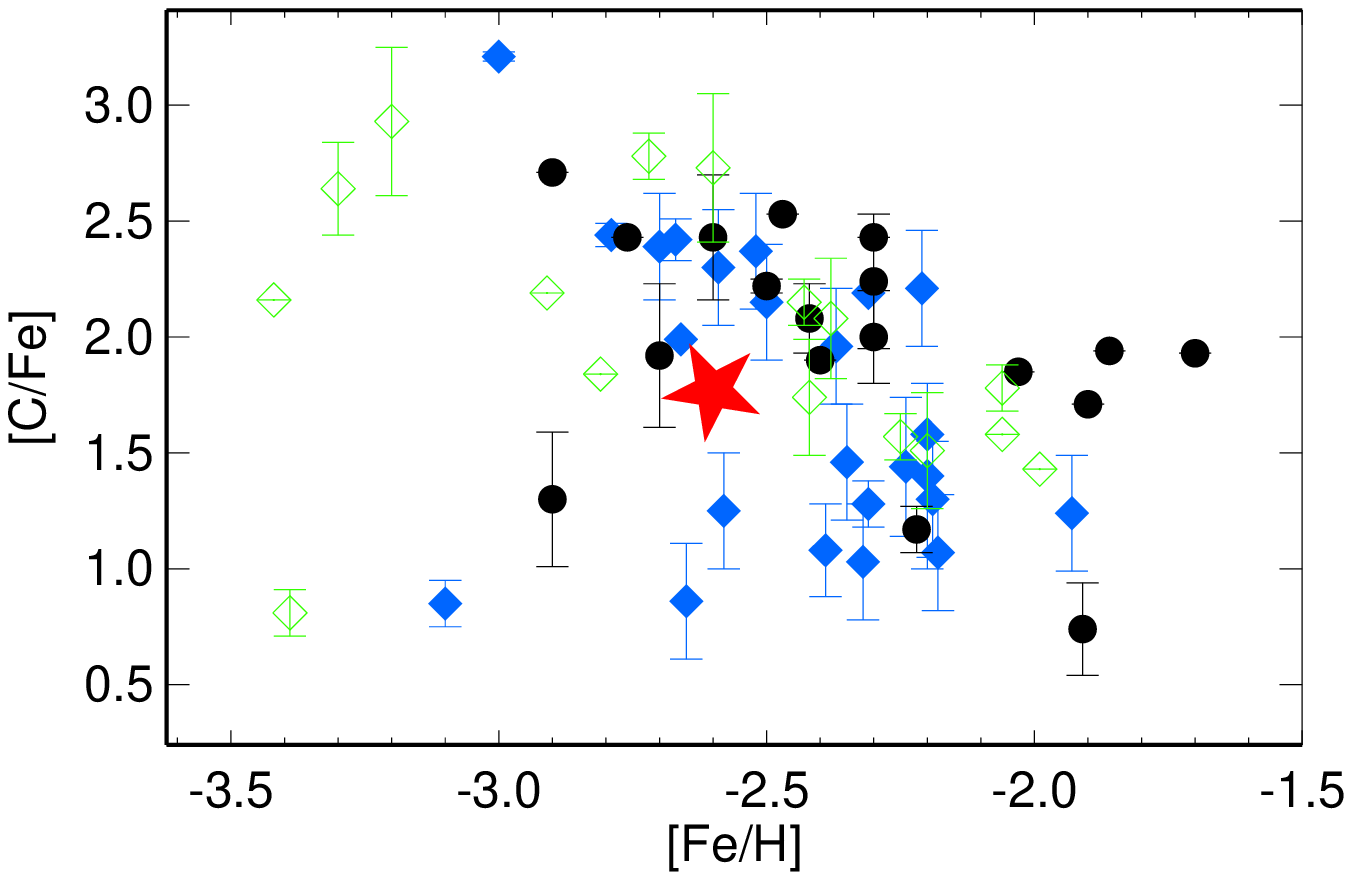}
\hspace{-4mm}	\includegraphics[width=0.33\textwidth, clip]{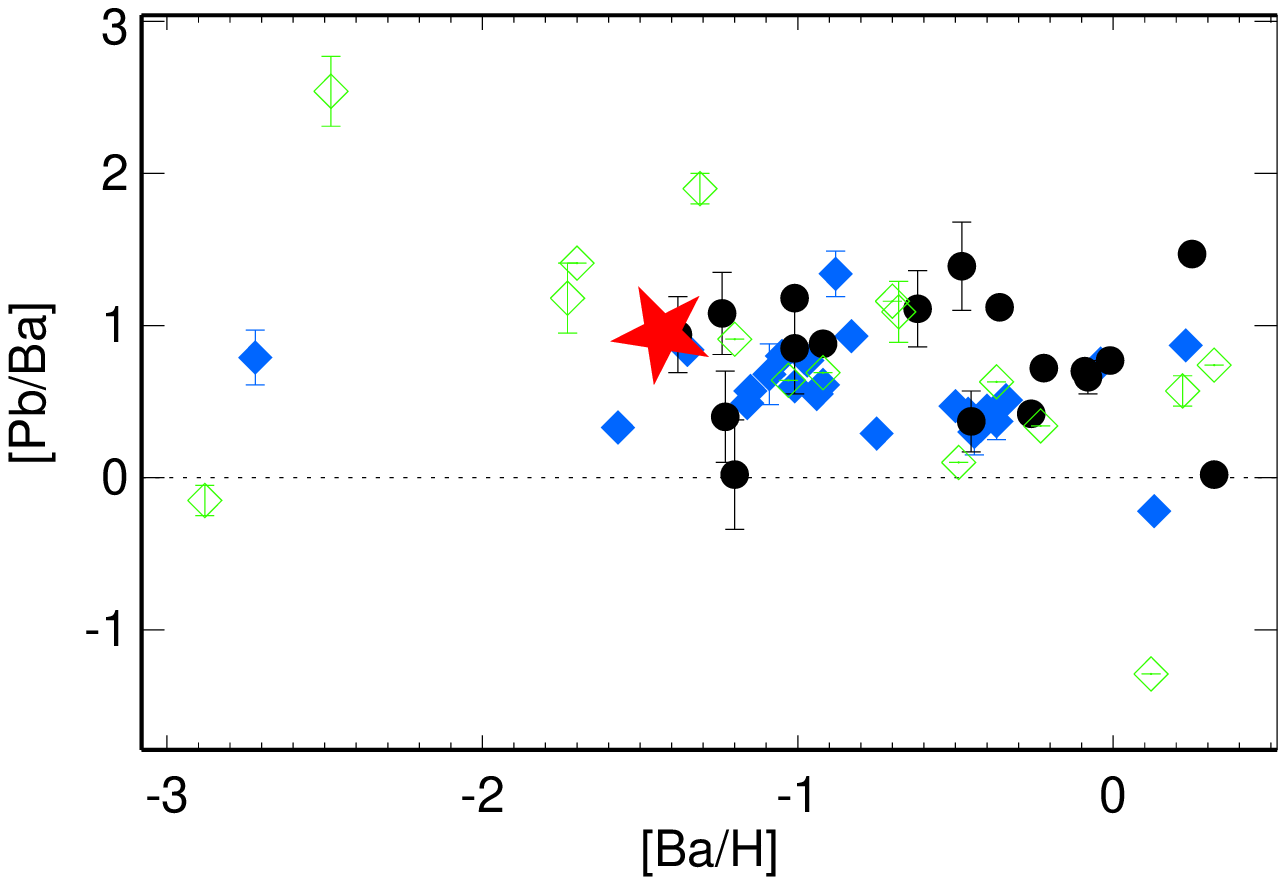}
\hspace{-4mm}	\includegraphics[width=0.33\textwidth, clip]{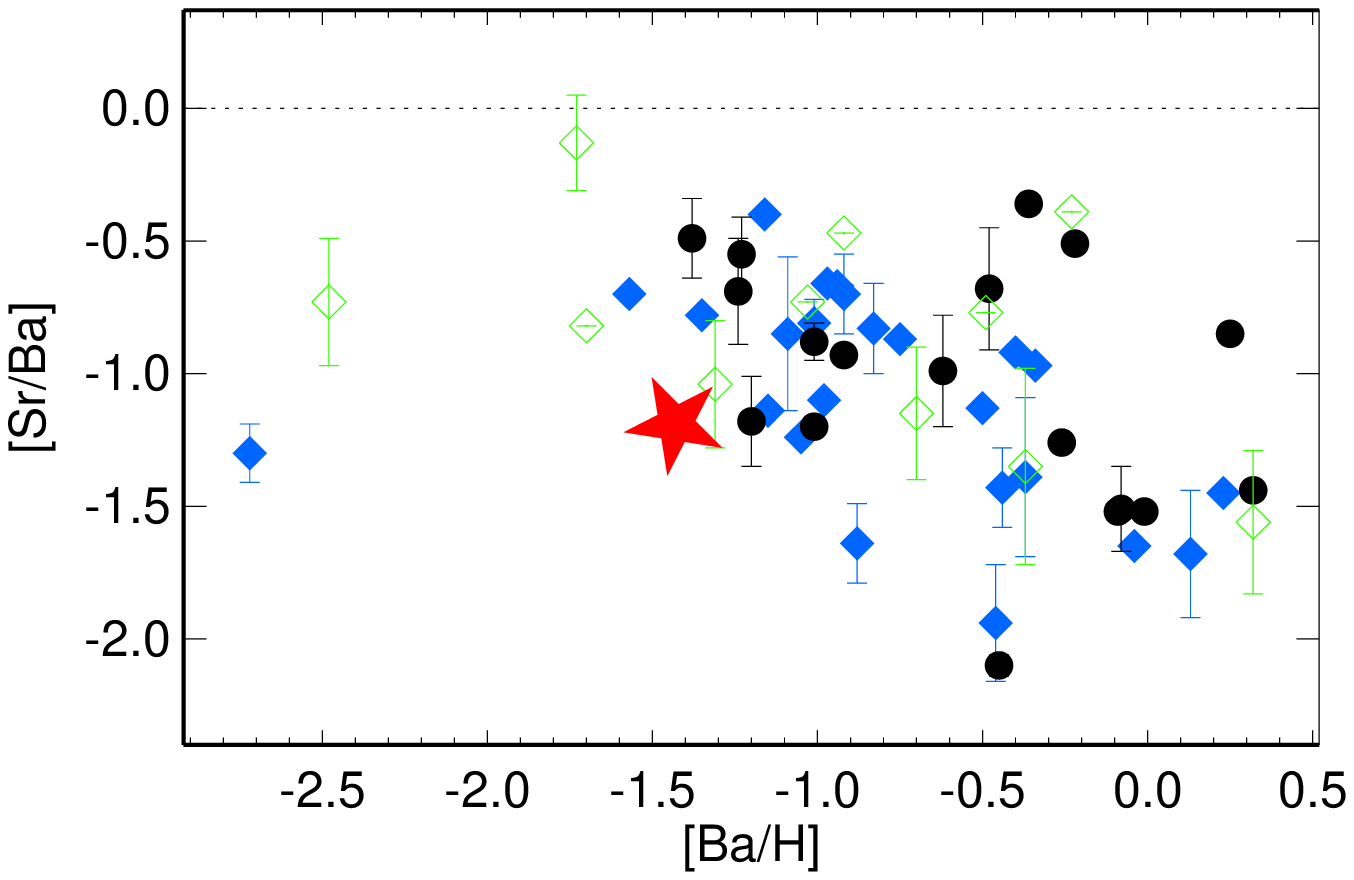}
  \caption{\label{Fig:cfe} CEMP-s (blue filled diamonds), CEMP-r/s (filled circles), and the stars with unknown Eu abundance (open green diamonds) in the abundance diagrams
  [C/Fe] versus [Fe/H] (left panel), [Pb/Ba] versus [Ba/H] (middle panel) and [Sr/Ba] versus [Ba/H].
 The data are taken from the SAGA database \citep{2008PASJ...60.1159S}. Pr184237 is shown by the red star symbol.}
\end{minipage}
\end{figure*}

In order to compare Pr184237 with the Galactic halo CEMP-s and CEMP-r/s stars, we used the Stellar Abundances for Galactic Archaeology (SAGA) database \citep{2008PASJ...60.1159S}. In total, 58 stars were selected with [Fe/H] $< -1.5$, [C/Fe] $> 0.7$, and the Pb abundance measured.
 We separated them into two subclasses, using the criteria by \citet{2005ARA&A..43..531B}, namely, [Ba/Fe] $> 1$ and [Ba/Eu] $> 0.5$ for CEMP-s and [Ba/Eu] = 0 to 0.5, [Ba/Fe] $> 0$ and [Eu/Fe] $> 0$ for CEMP-r/s.
 The obtained subsamples of 24 CEMP-s and 18 CEMP-r/s stars, as well as the remaining CEMP stars with unknown Eu abundance are displayed together with Pr184237 in Fig.~\ref{Fig:cfe}. The three groups of stars fully overlap on the abundance diagrams [Pb/Ba] and [Sr/Ba] versus [Ba/H]. 
 The position of Pr184237 in these diagrams does not contradict either the CEMP-s or the CEMP-r/s status. Only the lower [Ba/Eu] ratio compared to the criterion for the CEMP-s subclass leads us to refer to Pr184237 to as a CEMP-r/s star.

\section{Origin of the heavy element abundance pattern }\label{Sect:origin}

\begin{figure}  
 \begin{minipage}{85mm}
\centering
	\includegraphics[width=0.99\textwidth, clip]{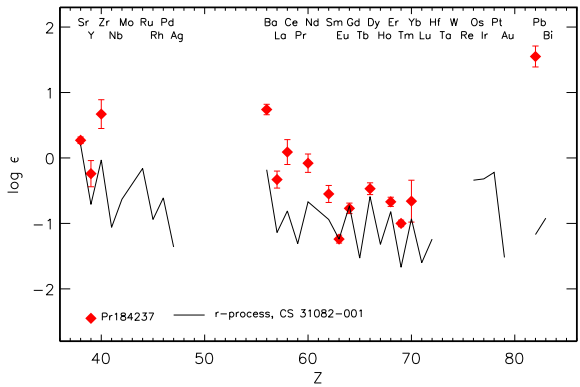}
  \caption{\label{Fig:rII} Heavy element abundance pattern of Pr184237 (diamonds) compared to abundances of an r-II star CS~31082-001 from \citet{2013A&A...550A.122S}. Abundances of CS~31082-001 were scaled to match the Eu abundance of Pr184237. }
\end{minipage}
\end{figure}

One commonly believes that the first heavy element nuclei in our Galaxy were produced in the r-process, although 
astrophysical sites for the r-process are still debated \citep[see][for a thorough review]{2021RvMP...93a5002C}. 
In the Galactic halo, a contribution of the main s-process to heavy element production becomes notable in stars starting from [Fe/H] $\simeq -2.6$, according to \citet{Simmerer2004}, or even from higher [Fe/H] $\simeq -1.4$ \citep{Roederer2010}. 

Spectroscopic analyses of MP stars in the inner Galaxy provide signatures of early r-process nucleosynthesis. \citet{2013ApJ...775L..27J} and \citet{2015Natur.527..484H,2016MNRAS.460..884H} report on europium enhancements at the level of [Eu/Fe] = 0.5 to 1.0.
Some of the stars, for example, 2MASS 18174532-3353235 \citep[${\rm [Fe/H] = -1.67}$, ${\rm [Eu/Fe]}$ = 0.99, ${\rm [Ba/Eu] = -0.5}$,][]{2013ApJ...775L..27J}  
and J181505.16-385514.9 \citep[${\rm [Fe/H] = -3.29}$, ${\rm [Eu/Fe]}$ = 0.96, ${\rm [Ba/Eu] = -0.92}$,][]{2015Natur.527..484H}, can be referred to as strongly r-process enhanced (r-II) stars, following the classification of \citet{2005ARA&A..43..531B}.
As shown by \citet{2012ApJ...749..175J,2016A&A...586A...1V}, and \citet{2019A&A...631A.113F,2022arXiv221005688F}, the r-process was the dominant neutron-capture production process in the  bulge until metallicity reached [Fe/H] $\sim -1$. 

 Therefore, we have every reason to believe that Pr184237 has formed from the matter where the  heavy elements were of the r-process origin. Later the star's atmosphere was enriched with the nuclear-processed material transferred from the evolved companion.
The enrichment scenario should explain
enhanced abundances of the heavy elements and also enhancements in C, N, O and the low $^{12}$C/$^{13}$C isotope ratio. 

\subsection{Pollution by the s-process products}\label{sect:rs}

In this scenario, we assume that heavy elements in Pr184237 originate from two sources. The first one is the interstellar matter out of which Pr184237 has formed, and it has the r-process element abundance pattern, with [r/Fe] = 1.03, as shown in Sect.~\ref{sect:heavy}. The second one is the s-process enriched material transferred onto a surface of Pr184237 from the more massive primary component when it evolved to the AGB phase. The transferred material was diluted with the atmospheric matter.

\begin{figure}  
 \begin{minipage}{85mm}
\centering
	\includegraphics[width=0.99\textwidth, clip]{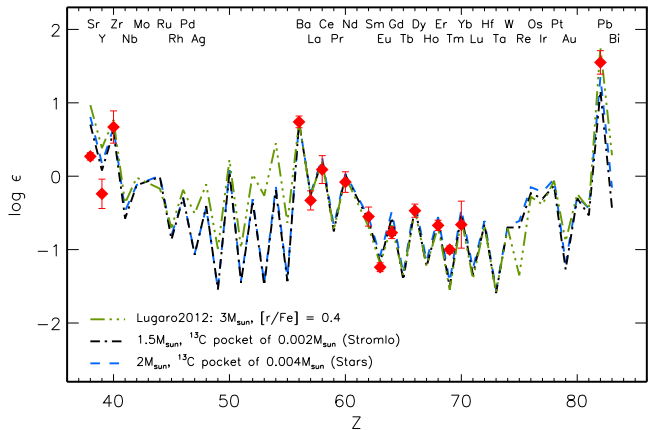}
  \caption{\label{Fig:s} Heavy element abundance pattern of Pr184237 (diamonds) compared to different s-process models from \citet{2012ApJ...747....2L}. See the text for more details. }
\end{minipage}
\end{figure}

Initial abundances of the heavy elements in Pr184237, $\varepsilon_{\rm r}$, were calculated using the empirical r-process relative yields, which were obtained from abundance analysis of a benchmark r-II star CS~31082-001 \citep{2013A&A...550A.122S}. The r-II stars reveal very similar heavy element abundance patterns \citep{2008ARA&A..46..241S}, suggesting a universal r-process.
In Fig.~\ref{Fig:rII}, the heavy element abundance pattern of Pr184237 is displayed together with the scaled abundances of CS~31082-001. The scaling factor was calculated from matching the Eu abundances of Pr184237 and CS~31082-001.
Abundances of Eu, Gd, Dy, and Er in Pr184237 lie well on the empirical r-process curve, suggesting an origin of these elements in the r-process. The Tm abundance, which is also expected to be on the r-process curve, appears to be higher. Three of the four inspected lines of Tm\ii, at 3700, 3701, and 3848~\AA, are heavily affected by noise in the spectrum of Pr184237, and the abundance from Tm\ii\ 3795~\AA\ seems to be overestimated. Neglecting Tm due to the uncertainty in its abundance, we computed the average [r/Fe] = 0.91$\pm$0.12.

For the s-process abundances, $\varepsilon_{\rm s}$, we used models of the s-process in the low-metallicity ([Fe/H] = $-2.3$) AGB stars computed by \citet{2012ApJ...747....2L} in the broad range of stellar masses (0.9--6 $M_{\odot}$) by varying the size of the $^{13}$C pocket, which is an uncertain parameter in AGB nucleosynthesis.

The quality of each enrichment scenario was judged in terms of the $\chi^2$ statistics. The statistical estimator is determined by

\begin{equation}
\chi^2 = \sum (\eps{obs} - \eps{mod})^2 / \sigma_{\rm obs}^2,
\label{formula2}
\end{equation}

\noindent where $\varepsilon_{\rm mod} = d\cdot \varepsilon_{\rm r} + (1 - d) \cdot \varepsilon_{\rm s}$, $d$ is a dilution factor, which is a free parameter and takes a value between 0 and 1,
 $\sigma_{\rm obs}$ is the abundance error, and the summation is carried out for the 15 elements from Sr to Pb observed in Pr184237.

\citet{2012ApJ...747....2L} provide two grids of the s-process abundances computed with two independent stellar evolution codes,
{\sc stars} and Stromlo. For the first grid, the best fit to the observations, with $\chi^2$ = 210 and $d$ = 0.982, was obtained using the model with an initial mass of $M_{init}$ = 2$M_{\odot}$ and the mass of the partial mixing zone $M_{mix}$ = 0.004$M_{\odot}$. Note that the $^{13}$C pocket forms in the partially mixed zone; we refer to \citet{2012ApJ...747....2L} for further details. In the Stromlo grid, the best model ($\chi^2$ = 200, $d$ = 0.980) has $M_{init}$ = 1.5$M_{\odot}$ and $M_{mix}$ = 0.002$M_{\odot}$. Both models are shown in Fig.~\ref{Fig:s} compared with the observations.

\citet{2012ApJ...747....2L} computed also the models with the assumption that an initial composition of the stars producing s-nuclei was different from a scaled-solar composition. Namely, the matter was r-process enriched with [r/Fe] = +0.4, +1.0, and +2.0. 
We neglect the models with [r/Fe] = +2.0, because the binary components should have common initial abundances, but Pr184237 does not reveal such a strong r-process enhancement. The other models are checked by adopting $\varepsilon_{\rm mod} = \varepsilon_{\rm s}$.
The best fit model ($\chi^2$ = 331) shown in Fig.~\ref{Fig:s} has $M_{init}$ = 3$M_{\odot}$, $M_{mix}$ = 0.0005$M_{\odot}$, and [r/Fe] = +0.4.

Judging by a purely numerical criterion, the Stromlo model of $M_{init}$ = 1.5$M_{\odot}$, with the least $\chi^2$ = 200, can be considered as the "best model". However, it overpredicts abundances of Sr and Y. This is in line with \citet{2012ApJ...747....2L}, who conclude that the light-s/heavy-s element abundance ratios in their models are too high compared with the observations of CEMP-r/s stars. 
 In contrast, the Pb abundance of the "best model" is lower compared with that for Pr184237. 
The observed Pb abundance is reproduced by the higher mass models, namely, of 2$M_{\odot}$ in the {\sc stars} grid and of 3$M_{\odot}$ in the grid of the r-process enhanced models.
Carbon enhancement of Pr184237 ([C/Fe] = 1.77) is reasonably well reproduced by each of three selected models ([C/Fe] = 1.51 to 1.94), when applying the dilution factor obtained in fitting the heavy element abundance pattern and assuming the initial C abundance to be equal to the scaled solar one. However, the predicted $^{12}$C/$^{13}$C isotope ratio is supersolar, in contrast to our finding for Pr184237.

\subsection{Pollution by the i-process products}\label{sec:iprocess} 


It is possible to reproduce the observed abundances of CEMP-r/s stars by using nucleosynthesis models for the intermediate (i) neutron capture process 
\citep{2014nic..confE.145D,2016ApJ...831..171H,2019ApJ...887...11H,2021A&A...649A..49G}. This nucleosynthesis
process introduced first by \citet{1977ApJ...212..149C} runs at neutron densities of $N_{\rm n}$ = 10$^{11}$-10$^{15}$~cm$^{-3}$, which are intermediate to that of the main s-process, which typically occurs at $N_{\rm n} \lesssim 10^{8}$~cm$^{-3}$, and considerably lower than r-process neutron densities. While He-shell burning in intermediate-mass AGB stars can result in $N_{\rm n} \sim 10^{12}$~cm$^{-3}$ \citep{2014ApJ...797...44F}, owing to the operation of the $^{22}$Mg($\alpha$,n)$^{25}$Mg reaction, the neutron exposure \citep{busso99} is likely not high enough to initiate an i-process. The operation of the i-process can produce enhancements in elements typically produced by both the s- and r-process (e.g., Ba and Eu), although the resulting abundance distribution is unique and not a superposition of the two \citep{2019ApJ...887...11H}.
Favourable conditions for the i-process can be found in multiple stellar sites, for example, proton-ingestion episodes (PIEs) in AGB stars, super-AGB stars, and rapidly accreting white dwarfs \citep[see, for example][]{2016ApJ...831..171H,2019ApJ...887...11H,2017ApJ...834L..10D,2019MNRAS.488.4258D,2021A&A...648A.119C,2022A&A...667A.155C}.

In order to understand an origin of the heavy elements in Pr184237, we explore the i-process nucleosynthesis in AGB stars, using the models of \citet{2019ApJ...887...11H} and \citet{2022A&A...667A.155C}, and in rapidly accreting white dwarfs (RAWD), using the models of \citet{2019MNRAS.488.4258D}. 
In the AGB star scenario, the investigated star is in a binary system and has accreted the i-process material from the AGB companion (which is now a white dwarf). In the scenario of enrichment of Pr184237 with the i-process products from the RAWD, the investigated star would have been a low-mass third companion orbiting the close, previously interacting, binary system with the RAWD. \citet{2019MNRAS.488.4258D} proposed that the i-process enriched star would have accreted a wind material ejected by the RAWD. There are two possible scenarios: either the RAWD exploded as a type Ia supernova (SN~Ia), and the i-process enriched star was ejected from the triple system, or it did not explode and the i-process enriched star is still part of the triple system. Variations in the radial velocity of Pr184237 suggest that our star could still orbit the surviving component in this scenario.

For each of the i-process models, we assume some dilution of the produced i-process nuclei with the heavy elements of the atmospheric matter, such that $\varepsilon_{\rm mod}$ in formula~(\ref{formula2}) is determined by

\begin{equation}
\log \varepsilon_{\rm mod} = \log (d\cdot \varepsilon_{\rm init} + (1 - d)\cdot\varepsilon_{\rm i-proc}),
\label{formula1}
\end{equation}

\noindent where $\varepsilon_{\rm i-proc}$ is the i-process abundance of a given element, $\varepsilon_{\rm init}$ is the initial atmospheric abundance, and $d$ is a dilution factor. The model abundances are fitted to the abundances of 15 heavy elements observed in Pr184237.

\begin{figure}  
 \begin{minipage}{85mm}
\centering
	\includegraphics[width=0.99\textwidth, clip]{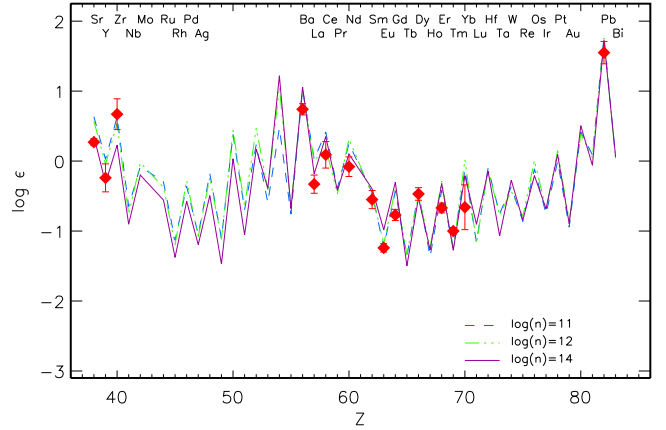}
  \caption{\label{Fig:hampel} Heavy element abundance pattern of Pr184237 (diamonds) compared to the best fit i-process models calculated following \citet{2019ApJ...887...11H}.}
\end{minipage}
\end{figure}


 In Fig.~\ref{Fig:hampel} we show the results from \citet{2019ApJ...887...11H} of an i-process calculated with constant neutron densities of different magnitude. The calculations assume conditions typical of the intershell region of a low-mass, low-metallicity AGB star. In the fitting procedure, the scaled solar abundances were adopted as $\varepsilon_{\rm init}$. The best three fits correspond to $N_{\rm n} = 10^{11}$~cm$^{-3}$ ($\chi^2 \simeq$ 107), 10$^{12}$~cm$^{-3}$ ($\chi^2 \simeq$ 115), and 10$^{14}$~cm$^{-3}$ ($\chi^2 \simeq$ 117), although qualitatively the fits look similar. Differences show up in the region between Sr and Ba, which are elements not observed for Pr184237.


\begin{figure}  
 \begin{minipage}{85mm}
\centering
	\includegraphics[width=0.99\textwidth, clip]{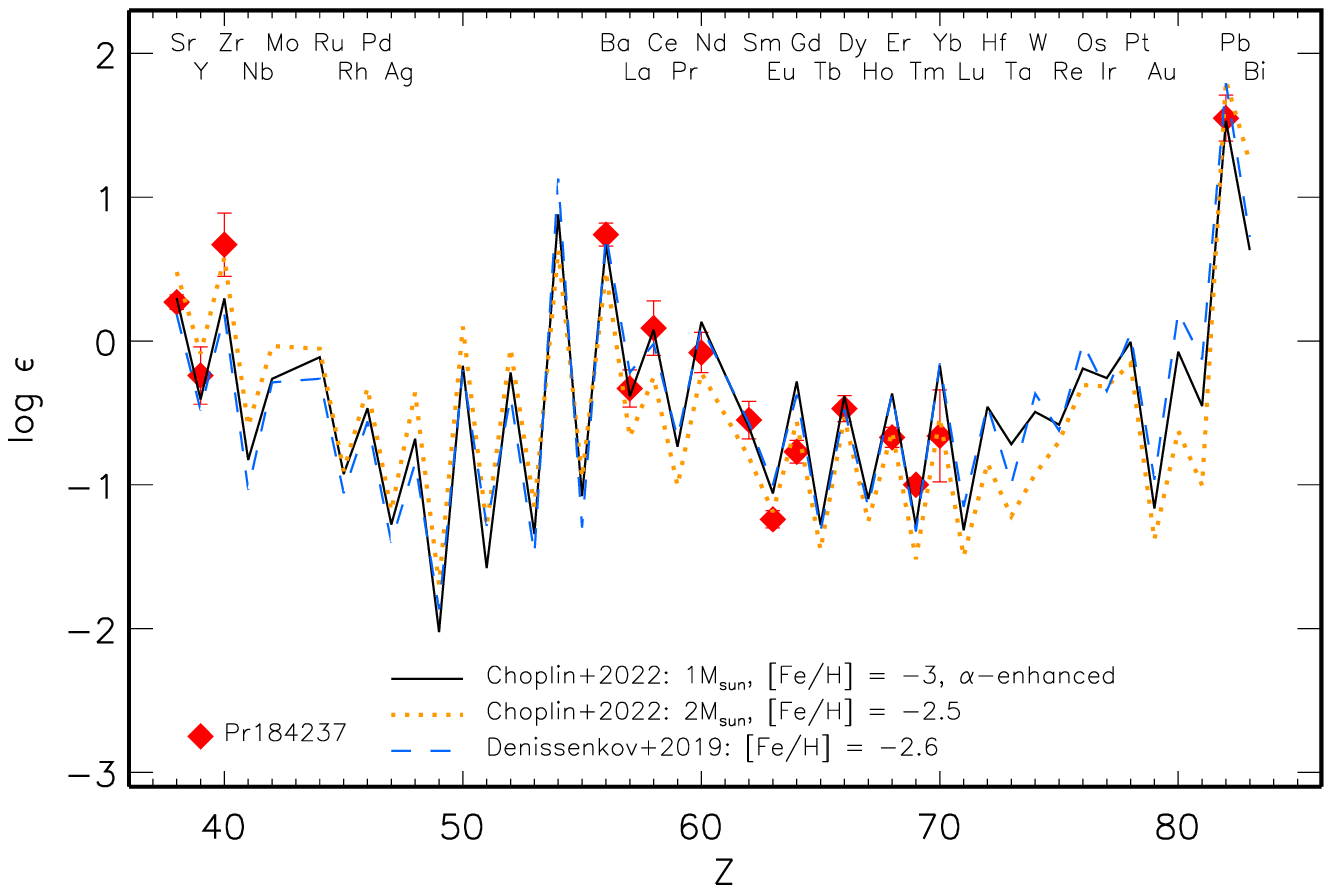}
  \caption{\label{Fig:choplin} Heavy element abundance pattern of Pr184237 (diamonds) compared to the models of an i-process nucleosynthesis in the AGB star \citep[][continuous and dotted curves for two models]{2022A&A...667A.155C} and the RAWD \citep[][dashed curve]{2019MNRAS.488.4258D}. See text for more details.}
\end{minipage}
\end{figure}


\citet{2022A&A...667A.155C} calculated 12 stellar evolution models with various initial masses and metallicities. Proton ingestion events happen in six of them, namely, in 1 and 2$M_{\odot}$ AGB models with [Fe/H] = $-3$, $-2.5$, and $-2.3$. The companion to Pr184237 had enough time to evolve to the AGB phase. \citet[][Table~1]{2022A&A...667A.155C} predicted total lifetimes of 6.3 and 0.8~Gyr for the $M$ = 1 and 2~$M_\odot$ models, respectively. When fitting the i-process yields to the observed heavy element abundance pattern, we assume that they are diluted with $\varepsilon_{\rm init}$ = $\varepsilon_{\rm r}$. Here, $\varepsilon_{\rm r}$ is determined as in the previous subsection.
 We obtain that the model with $M_{init}$ = 1$M_{\odot}$, [Fe/H] = $-3$, and $\alpha$-enhancement (M1.0z3.0a)
 has the smallest $\chi^2$ = 109 ($d$ = 0.98) and reproduces well the heavy element abundance pattern of Pr184237, including the light neutron-capture elements Sr and Y (Fig.~\ref{Fig:choplin}).
There are a few subtle differences though for the r-process elements Gd, Dy, Er and also for Zr and Yb. 
 This model predicts two pulses, and the PIE arises during the second one, such that the AGB surface is enriched in i-process products right after the PIE. With the dilution factor $d$ = 0.98 and the scaled solar abundance as the initial C abundance, the M1.0z3.0a  model predicts the present C enhancement as [C/Fe] = 1.42, which is slightly less than the observed value, and the isotope ratio of $^{12}$C/$^{13}$C = 4.3, which is close to the observed value.

Figure~\ref{Fig:choplin} also shows the model for 2$M_{\odot}$ and [Fe/H] = $-2.5$ (M2.0z2.5). Keeping in mind that the Tm abundance of Pr184237 seems to be overestimated (see Sect.~\ref{sect:rs}), we find the best fits to the observed abundance pattern that does not include Tm. The lowest $\chi^2$ = 54.7 ($d$ = 0.95) was obtained for the M2.0z2.5 model. However, the M1.0z3.0a model has a similarly small $\chi^2$ = 58.4 ($d$ = 0.985). Compared to M1.0z3.0a, the M2.0z2.5 model reproduces well observed abundances of Eu to Yb (Tm is not included) and not only Sr and Y, but also Zr. However, M2.0z2.5 is worse for Ba, La, Ce and, with $^{12}$C/$^{13}$C = 24.5, also for the carbon isotope ratio. Right after the PIE pulse the model predicts low $^{12}$C/$^{13}$C = 3.5, but an enrichment of the AGB surface in $^{13}$C reduces during ten subsequent thermal pulses due to a dilution of the i-process material. 



 For an i-process in RAWD, we use the G (iRAWD\_2.6) model of \citet{2019MNRAS.488.4258D}  with exactly the same metallicity as [Fe/H] = $-2.6$ for Pr184237. They calculated an evolution of a close binary system with $M_1$ = 2.5$M_{\odot}$ and $M_2$ = 2.0$M_{\odot}$ and the i-process nucleosynthesis when the primary reached the stage of rapidly accreting white dwarf. The best fit ($\chi^2$ = 119) displayed in Fig.~\ref{Fig:choplin} was obtained with a dilution factor of $d$ = 0.9997.  
 The RAWD model predicts slightly greater [C/Fe] = 2.02 value than the observed C enhancement, but extremely high $^{12}$C/$^{13}$C isotope ratio.

\subsection{Uncertainties in the choice of enrichment scenario}\label{sect:uncert}

 In all of the considered scenarios, the initial atmospheric chemical composition of Pr184237 is polluted by nucleosynthesis products transferred from the more massive, evolved companion. Can we decide on the type of nucleosynthesis and initial mass of the companion from a comparison of our observed data with the theoretical models?

Compared with an i-process, we have more arguments against the s-process enrichment. The best fit s-process models have greater $\chi^2$, by a factor of more than two. This is mostly due to the overproduction of the light neutron-capture elements Sr and Y. When ignoring Sr to Zr, we obtained $\chi^2$ = 119 for the Stromlo model of 1.5$M_{\odot}$ shown in Fig.~\ref{Fig:s}. The s-process models produce a C enhancement close to the observed one, but with much lower enhancements in N and O. For example, assuming the initial abundances [N/Fe] = 0 and [O/Fe] = 0.5, we obtain [N/Fe] = 0.14 and [O/Fe] = 0.68 after the pollution. To remind the reader, Pr184237 reveals [N/Fe] = 1.29 and [O/Fe] = 1.32. The s-process models predict very high $^{12}$C/$^{13}$C isotope ratios, however, this and the low [N/Fe] predicted from the model can be due to uncertainties in the theoretical models of low-metallicity AGB stars. 

 Of two sites for an i-process, AGB and RAWD, no preference can be given to any of them with respect to $\chi^2$ and enhancements in C, N, and O. For Pr184237, we derive supersolar and similar [C/N] and [C/O]  ratios at the level of 0.45. Supersolar [C/O] $\simeq$ 0.3 ratios are predicted by the M1.0z3.0a and M2.0z2.5 models of \citet{2022A&A...667A.155C} and the iRAWD\_2.6 model of \citet{2019MNRAS.488.4258D}. The initial O abundance was assumed to be [O/Fe] = 0.5. There are discrepancies in [C/N] between the models and observations, however, we realise that our estimate of the N abundance in Pr184237 is less reliable compared with that for C and O.

From a chemistry point of view, an advantage of the AGB i-process models of \citet{2022A&A...667A.155C} is a subsolar $^{12}$C/$^{13}$C ratio at the surface of the polluted star which tends to be in line with observations of Pr184237.

We realise that our conclusion about a preferrable enrichment scenario can be influenced by the uncertainties in the theoretical models and in the element abundances determined for Pr184237.

\subsubsection{Uncertainties in modelling the i-process}

 Theoretical modelling of low and intermediate-mass stars on the AGB comes with significant uncertainties when considering mixing between the He-shell and envelope, and from mass-loss, some of which is likely accreted onto the companion by stellar wind accretion. For low-metallicity AGB stars the situation is even more complicated owing to proton-ingestion episodes, which can drive an i-process, along with the previous uncertainties \citep[we refer to][for a detailed discussion]{2014PASA...31...30K}.
\citet{2021A&A...648A.119C} perform detailed AGB calculations of proton-ingestion and found that the event can terminate the AGB phase, a phenomenon not found by \citet{2008A&A...490..769C}.
However Choplin~et~al. have coupled the surface opacity to the envelope enrichment, which affects the structure significantly after the star becomes carbon rich \citep[e.g., see also][]{2002A&A...387..507M}.
These authors also examine the effect of modelling uncertainties on the nucleosynthesis predictions and note an uncertainty of $\pm 0.3$~dex in the abundances. This value does not take uncertainties from nuclear physics into consideration, where many nuclei are unstable and consequently have uncertain neutron capture cross sections. 

An impact of the uncertainties in the (n,$\gamma$) reaction rates and $\beta$-decay rates for 164 unstable isotopes, from $^{131}$I to $^{189}$Hf, on the predicted i-process abundances was investigated by \citet{2021MNRAS.503.3913D} with their RAWD ([Fe/H] = $-2.6$) model for the elements from Ba to W. \citet{2021MNRAS.503.3913D} show that variations in the (n,$\gamma$) reaction rates result in notable changes of up to 0.5~dex (see their Figs.~7, 10, 15) in the predicted abundances.

For the RAWD model from \citet{2019MNRAS.488.4258D}, the main uncertainty is that the mass accretion rate is held steady at a low value, which leads to H-flashes that cause the envelope of the accreted matter to expand. In order to stop the model from becoming a red giant, an artificial mass-loss is assumed to bring the model back to a steady mass accretion. Secondly, the low-metallicity WDs tend to grow significantly, nearing the Chandrasekhar limiting mass and may explode as a type Ia supernova. While low-metallicity SNeIa likely occurred, they must have been very rare in the early Universe or we would see evidence of this in the chemical record (e.g., lower [$\alpha$/Fe], higher Mn abundance, etc). 
\citet{2018ApJ...854..105C} studied the i-process contribution from RAWDs to the chemical enrichment of the galaxy, and included an exploration of the number of RAWD systems using binary population synthesis. Interestingly in the binary population synthesis model, the systems that evolved to become RAWDs avoided SN~Ia altogether, as a consequence of artificially suppressing H accumulation onto the WD. While this may avoid the issues discussed above in regard to low metallicity SN~Ia, it also suggests that the WD growth seen in the models of \citet{2019MNRAS.488.4258D} may not happen.

In summary, the i-process may have occurred in multiple stellar sites in the early Galaxy including 
low-metallicity AGB stars, rapidly rotating white dwarfs, and even perhaps massive stars \citep{2018ApJ...865..120B,2018MNRAS.474L..37C}. It is unclear as yet which of these sites polluted the stars with neutron-capture elements, with all sites still showing considerable uncertainties stemming from modelling and input physics.


\subsubsection{Uncertainties in the abundance determinations}

As already noted, we probably overestimate the Tm abundance. If 
we remove Tm from the the heavy element abundance pattern, the AGB star of not only $M_{\rm init}$ = 1.0, but also $M_{\rm init}$ = 2$M_{\odot}$ can be the site of an i-process that enriched Pr184237.
For the s-process models, removing Tm decreases $\chi^2$, for example to $\chi^2$ = 93.7 for the 1.5$M_{\odot}$ shown in Fig.~\ref{Fig:s}, but the best fit models remain the same.

 The analysis of the enrichment scenarios was based on the LTE abundances of Pr184237. When using the NLTE abundances for Sr, Ba, Eu, and Pb and the LTE abundances for the remaining elements, we obtain slightly different $\chi^2$ = 140 ($d$ = 0.99995) for the iRAWD\_2.6 model. In the case of the AGB i-process, the best fit ($\chi^2$ = 110, $d$ = 0.99) is achieved for the same mass (1.0$M_{\odot}$), but a higher metallicity of [Fe/H] = $-2.5$. The best fits for the s-process models in the Stromlo grid are shifted to the higher masses of 2 and 2.5$M_{\odot}$. We also made fitting the i-process models of \citet{2022A&A...667A.155C} to the NLTE abundances of Sr, Ba, Eu, and Pb without including the other heavy elements and obtained the same solution for the best fit model, that is M1.0z2.5. NLTE leads to a 0.66~dex higher Pb abundance of Pr184237 compared with the LTE value, and none of the considered models predicts such a high Pb abundance. The difference between the observed and theoretical values amounts to 0.3~dex for the \citet{2022A&A...667A.155C} model and exceeds 0.5~dex for the iRAWD\_2.6 and the s-process models.
 
When fitting the i-process models of \citet{2022A&A...667A.155C} and \citet{2019MNRAS.488.4258D}, we assume that the initial chemical composition has the r-process pattern. We tested whether using the scaled solar abundances, $\varepsilon_{\odot}$, can influence on the stellar parameters of the best fits. The present solar system abundances were taken from \citet{2021SSRv..217...44L}.
The obtained results are very similar to that for $\varepsilon_{\rm init}$ = $\varepsilon_{\rm r}$. The least $\chi^2$ = 116 ($d$ = 0.97) was achieved for the same AGB model of 1$M_{\odot}$ and [Fe/H] = $-3$. The $\chi^2$ slightly increases up to $\chi^2$ = 122 ($d$ = 0.9997) for the iRAWD\_2.6 model.

 Thus, an i-process nucleosynthesis in the AGB stars with a progenitor mass of 1.0-2.0$M_{\odot}$, as calculated by \citet{2022A&A...667A.155C}, is a promising  scenario to explain the CEMP-r/s phenomenon at the surface of Pr184237. 




\section{Conclusions}\label{Sect:conclusion}

In a sample of 20 VMP stars selected from the PIGS survey for a high-resolution spectroscopic follow-up with UVES/VLT2, a FERRE analysis identified a CEMP star, Pr184237, which is strongly enhanced also in Ba, La, and Eu.  Its orbit is confined to within $\sim 2.6$~kpc of the Galactic centre, and it is most likely an early Milky Way star rather than accreted. In this study, we improved the star's atmospheric parameters: $\Teff$ = 5100~K, $\logg$ = 2.0, and [Fe/H] = $-2.60$ and determined abundances of 32 chemical elements, including 15 heavy elements beyond the iron group. The atmospheric parameters and abundances of 13 elements, from Na to Pb, were derived based on the NLTE line formation.

We improved the C abundance, [C/Fe] = 1.77, and found that Pr184237 is also enhanced in N and O. The star reveals a low carbon isotope ratio of $^{12}$C/$^{13}$C = 7, which is a signature of nuclear-processed material. 
 Pr184237 is far from the AGB evolutionary phase, and the self-pollution hypothesis should be rejected.
The element abundance pattern in the Na-Zn range is typical of the Galactic halo stars. 
However, the neutron-capture elements, in particular lead with [Pb/Fe] = 2.12 (LTE) and 2.78 (NLTE), are strongly enhanced.
With [Ba/Eu] = 0.32, Pr184237 is the first star of the CEMP-r/s subclass in the inner Galaxy. Compared with the halo CEMP-r/s stars, Pr184237 reveals very similar abundance ratios among the representative elements of the first, second and third s-process peaks, namely, Sr, Ba and Pb. 

We considered two scenarios of chemical enrichment for Pr184237. In both cases, the star is likely in a binary or higher order multiple system, consistent with the detected radial velocity variability.

In the first scenario, the star would have been born from the r-process rich material enriched by previous stellar generations, and would have been polluted with the s-process elements by a binary companion (former AGB) -- this is the r+s scenario, where the different heavy elements come from different sources. We applied the empirical r-process yields based on abundances of an r-II star CS~31082-001 \citep{2013A&A...550A.122S} and the s-process abundances from calculations of metal-poor AGB stars by \citet{2012ApJ...747....2L} and fitted a combination of them. We find that some of the observed abundances of the first s-process peak elements (Sr, Y) cannot be explained in the r + s scenario. This scenario does not reproduce the subsolar $^{12}$C/$^{13}$C isotope ratio derived for Pr184237. 

 In the second scenario, the initial chemical composition could be polluted with both s- and r-process elements produced by the i-process nucleosynthesis, which can occur in the companion AGB star or rapidly accreting white dwarf. We checked the i-process models from \citet{2019ApJ...887...11H}, \citet{2022A&A...667A.155C} -- both for AGB stars and \citet[][for RAWDs]{2019MNRAS.488.4258D}. For each grid, the best fit models reproduce reasonably well the heavy element abundance pattern of Pr184237. The models from \citet{2022A&A...667A.155C} and \citet{2019MNRAS.488.4258D} reproduce also the C enhancement and the C/O ratio. A subsolar $^{12}$C/$^{13}$C isotope ratio is only obtained when using the AGB models of 1.0-2.0$M_{\odot}$ from calculations of \citet{2022A&A...667A.155C}. 

  This study supports the hypothesis that the CEMP-r/s stars, like all CEMP-s stars, form in binary or higher order multiple system \citep[see][and references therein]{2016A&A...587A..50A,2021A&A...649A..49G}.
We consider an i-process nucleosynthesis in the AGB stars with a progenitor mass of 1.0-2.0$M_{\odot}$ and a transfer of the nuclear-processed material onto the companion surface as a promising scenario for explaining the CEMP-r/s phenomenon of Pr184237. This suggests that the i-process could contribute to the chemical evolution of carbon and heavy elements in the inner Galaxy.




\section*{Acknowledgements}
This study is based on observations made with the Very Large Telescope of the European South Observatory (Chile) under program ID 0105.B-0078(A). This research has made use of the data from the European Space Agency (ESA) mission Gaia\footnote{\url{https://www.cosmos.esa.int/gaia}}, processed by the Gaia Data Processing and Analysis Consortium (DPAC\footnote{\url{https://www.cosmos.esa.int/web/gaia/dpac/consortium}}). Funding for the DPAC has been provided by national institutions, in particular the institutions participating in the Gaia Multilateral Agreement.
This research has made use of the data from the VALD, SAGA, and ADS\footnote{\url{http://adsabs.harvard.edu/abstract\_service.html}} databases.
This research was enabled in part by support provided by BC DRI (British Columbia Digital Research Infrastructure) Group and the Digital Research Alliance of Canada\footnote{\url{https://alliancecan.ca}}.
AA acknowledges support from the Herchel Smith Fellowship at the University of Cambridge and a Fitzwilliam College research fellowship supported by the Isaac Newton Trust. 
 DA acknowledges financial support from the Spanish Ministry of Science and Innovation (MICINN) under the 2021 Ram\'on y Cajal program MICINN RYC2021‐032609. 
 AIK was supported by the Australian Research Council Centre of Excellence for All Sky Astrophysics in 3 Dimensions (ASTRO 3D), through project number CE170100013. 
 NFM gratefully acknowledges support from the French National Research Agency (ANR) funded project ``Pristine'' (ANR-18-CE31-0017) along with funding from the European Research Council (ERC) under the European Unions Horizon 2020 research and innovation programme (grant agreement No. 834148). 
 FS thanks the Dr. Margaret "Marmie" Perkins Hess postdoctoral fellowship for funding his work at the University of Victoria.
 JIGH acknowledges financial support from the Spanish Ministry of Science and Innovation (MICINN) project PID2020-117493GB-I00.
 
\section{Data availability}

All our main results are available in the tables in this article. The UVES spectrum is publicly available on the ESO archive. Other data underlying this article will be shared on reasonable request to the corresponding author.

\bibliography{pr184237,mashonkina,nlte,references,atomic_data}
\bibliographystyle{mnras}

\label{lastpage}
\end{document}